\documentclass[12pt]{article}
\usepackage{hyperref}
\usepackage{amsmath}
\usepackage{amssymb,amsthm}
\usepackage[english]{babel}
\usepackage[textwidth=18cm,textheight=22cm]{geometry}
\usepackage{array}
\newcommand{\C}{\mathbb{C}}
\newcommand{\R}{\mathbb{R}}
\newcommand{\Z}{\mathbb{Z}}
\newcommand{\g}{\mathfrak{g}}

\renewcommand{\d}{\mathrm{d}}
\newcommand{\bt}{\boldsymbol{t}}

\newcommand{\bu}{\boldsymbol{U}}

\newtheorem{pro}{Proposition}

\newcommand{\Exp}[1]{\operatorname{e}^{#1}}

\begin{document}

\title{$S$-functions, reductions and hodograph solutions of the
\\$r$-th dispersionless modified KP and Dym hierarchies}

\author{Manuel Ma{\~n}as\\ Departamento de F{\'\i}sica Te{\'o}rica II, Universidad Complutense\\ 28040-Madrid, Spain\\
email: manuel@darboux.fis.ucm.es}

\maketitle

\abstract{We introduce an $S$-function formulation for the
recently found $r$-th dispersionless modified KP and $r$-th
dispersionless Dym hierarchies, giving also a connection of these
$S$-functions with the Orlov functions of the hierarchies. Then,
we discuss a reduction scheme for the hierarchies that together
with the $S$-function formulation leads to hodograph systems for
the associated solutions. We consider also the connection of these
reductions with those of the dispersionless KP hierarchy and with
hydrodynamic type systems. In particular, for the 1-component and
2-component reduction we derive, for both hierarchies, ample sets
of examples of explicit solutions.}

\section{Introduction}

Dispersionless integrable hierarchies, which  is an active line of
research in the theory of integrable systems, originates from
several sources. Let us metion mention here the pioneering work of
Kodama and Gibbons \cite{kodama,kodama-gibbons} on the
dispersionless KP, of Kupershimdt on the dispersionless modified
KP \cite{ku} and the important work of Tsarev  on the role of
Riemann invariants and hodograph transformations \cite{tsarev}.
Another approach is that of Takasaki and Takabe,
\cite{takasaki-takebe}, \cite{takasaki-takebe_toda} and
\cite{takasaki-takebe-3} which gave the Lax formalism, additional
symmetries, twistor formulation of the dispersionless KP and
dispersionless Toda hierarchies.

It is also worthwhile mentioning the role of dispersionless
systems in topological field theories, see \cite{krichever} and
\cite{dubrovin}. More recent progress appears in relation with the
theory of conformal maps \cite{gibbons} and \cite{wiegman},
quasiconformal maps and $\bar\partial$- formulation
\cite{konopelchenko},  additional symmetries \cite{martinez-manas}
and twistor equations \cite{Previo}, on hodograph equations for
the Boyer--Finley equation \cite{works_on_boyer} and its
applications in General Relativity, see also \cite{dunajski}.  An
new and interesting approach is presented in \cite{ferapontov}.
Finally, we remark the contribution on geometrical optics and the
dispersionless Veselov--Novikov equation
\cite{konopelchenko-moro}.

Recently, a new Poisson bracket and associated Lie algebra
splitting was presented in \cite{blaszak} to construct new
dispersionless integrable hierarchies, as  the $r$-th modified
 dispersionless KP hierarchy ($r$-dmKP) and  the $r$-th
 dispersionless Dym hierarchy ($r$-dDym), and latter on, see
\cite{blaszak2}, the theory was further extended. Moreover, we
studied in \cite{manas} the factorization of canonical
transformations in these Poisson algebras to get a new hierarchy,
the $r$-th dispersionless Toda ($r$-dToda) hierarchy which
contains the $r$-dmKP and $r$-dDym hierarchies as particular
cases. For this new hierarchy  we found additional symmetries and
a new Miura map among the $r$-dmKP and the $r$-dDym hierarchies.

In this paper we extend our results on the theory of reductions of
the dispersionless KP (dKP) hierarchy \cite{otros} and of the
Whitham hierarchies \cite{guil-manas-martinez}, to the $r$-dmKP
and $r$-dDym hierarchies introduced in \cite{blaszak}. The
keystone is the $S$-function formulation of the integrable
hierarchies and a reduction scheme that leads to hodograph systems
characterizing solutions.

 The layout of the paper is as follows. In \S 2 we briefly present
 the $r$-dmKP and $r$-dDym hierarchies. Then, in \S 3 we discuss
 the $S$ functions for these hierarchies, firstly we present the
 $S$-functions as potentials of the hierarchies, then we analyze
 their relation with the Orlov functions we introduced for these
 hierarchies in \cite{manas} and finally we reformulate the
 integrable hierarchies in terms of $S$-functions. To end, in \S
 4, we extend our results on the dKP and Whitham hierarchies to the
 $r$-dmKP and $r$-dDym hierarchies. For each hierarchy we discuss
 the reductions scheme, the associated hydrodynamic type systems,
 hodograph systems and explicit examples of
 solutions for the 1-component and 2-component reductions.

 We must underline that the additional symmetries found in
 \cite{manas} may be applied to the described solutions getting in this
 manner even more general sets of solutions.

\section{The integrable hierarchies}

The hierarchies we shall consider in this article were  introduced
in \cite{blaszak} within the Lax formalism as follows. The setting
needs of the Lie algebra $\g$ of Laurent series
$H(p,x):=\sum_{n\in\Z}u_n(x)p^n$ in the variable $p\in\C$ with
coefficients depending on the variable $x\in\R$, with Lie
commutator given by the following Poisson bracket
\begin{gather}\label{bracket}
\{H_1,H_2\}=p^r\Big(\frac{\partial H_1}{\partial p}\frac{\partial
H_2}{\partial x}-\frac{\partial H_1}{\partial x}\frac{\partial
H_2}{\partial p}\Big),\quad r\in\Z.
\end{gather}
We shall use the following triangular type splitting of $\g$ into
Lie subalgebras
\begin{equation}\label{splitting}
\g=\g_>\oplus\g_{1-r}\oplus \g_<
\end{equation}
where
\begin{equation*}
\g_\gtrless:=\C\{u_n(x)p^n\}_{n\gtrless(1-r)},\quad
\g_{1-r}:=\C\{u(x)p^{1-r}\},
\end{equation*}
and therefore fulfil  the following property
\[
\{\g_\gtrless,\g_{1-r}\}= \g_\gtrless.
\]

\subsection{The $r$-th dispersionless modified KP hierarchy}
If we define the Lie subalgebra $\g_\geqslant$ as
\[
\g_\geqslant:=\g_{1-r}\oplus\g_>
\]
we have  the  direct sum decomposition of the Lie algebra $\g$
given by
\begin{equation}\label{algebra-factor}
\g=\g_<\oplus\g_\geqslant,
 \end{equation}
 and the associated resolution of the identity into projectors
 \[
 1=P_\geqslant+P_<.
 \]

Given  a Lax function $L$ of the form
\begin{gather}\label{lax.func}
L=p+u_0(x)+u_1(x)p^{-1}+u_2(x)p^{-2}+\cdots,\quad p\to\infty,
\end{gather}
we introduce
\begin{align}\label{omega}
  \Omega_n:=P_\geqslant L^{n+1-r},\quad n=1,2,\dots.
\end{align}

The $r$-dmKP hierachy is the following set of Lax equations
\begin{gather}\label{Lax.eqs}
\frac{\partial L}{\partial t_n}=\{\Omega_n,L\},\quad n=1,2,\dots,
\end{gather}
where we have introduced an infinite set of time variables
$\{t_n\}_{n=1}^\infty$. 
 One easily deduce that the
first equations of this hierarchies are
\begin{gather}\label{dmkp-system}
\begin{aligned}
u_{0,t_1}&=(2-r)u_{1,x}+(2-r)(1-r)u_0u_{0,x},\\
u_{1,t_1}&=(2-r)u_{2,x}+(2-r)(1-r)u_0u_{1,x}+(2-r)u_1u_{0,x},\\
u_{0,t_2}&=(3-r)u_{2,x}+(3-r)(2-r)(u_0u_{1,x}+u_{0,x}u_1)+\frac{1}{2}(3-r)(2-r)(1-r)u_0^2u_{0,x}.
\end{aligned}
\end{gather}
This system, once $u_1$ and $u_2$ are expressed in terms of $u_0$
leads to the $r$-dmKP equation for $u:=u_0$
\begin{multline}\label{r-dmKP}
u_{t_2}=\frac{3-r}{(2-r)^2}(\partial_x^{-1}u)_{t_1t_1}+
\frac{(3-r)(1-r)}{2-r}u_x(\partial_x^{-1}u)_{t_1}+\frac{r(3-r)}{2-r}uu_{t_1}-
\frac{(3-r)(1-r)}{2}u^2u_x.
\end{multline}

\subsection{The $r$-th dispersionless Dym hierarchy}
We now take the Lie subalgebra $\g_\leqslant$ as
\[
\g_\leqslant:=\g_{1-r}\oplus\g_<
\]
and consider
\begin{equation}\label{algebra-factor}
\g=\g_>\oplus\g_\leqslant,
 \end{equation}
 and the corresponding resolution of the identity into projectors
 \[
 1=P_>+P_\leqslant.
 \]
Given the Lax function $\tilde L$ as follows
\begin{gather}\label{llax.func}
\tilde L= v p+v_0(x)+v_1(x)p^{-1}+\cdots,\quad p\to\infty,
\end{gather}
we introduce
\begin{align}\label{tilde.omega}
  \tilde \Omega_n:=P_> \tilde L^{n+1-r}, \quad n=1,2,\dots
\end{align}
The $r$-dDym hierarchy is defined by
\begin{gather}\label{LLax.eqs}
\frac{\partial \tilde L}{\partial t_n}=\{\tilde \Omega_n,\tilde
L\},\quad n=1,2,\dots.
\end{gather}
The first equations of this hierarchy are
\begin{gather}\label{dDym-system}
\begin{aligned}
v_{t_1}&=(2-r)v^{2-r}v_{0,x},\\
v_{0,t_1}&=(2-r)v^{1-r}(vv_1)_x,\\
v_{t_2}&=(3-r)v^{2-r}(vv_1)_x+(3-r)(2-r)v^{2-r}v_0v_{0,x}.
\end{aligned}
\end{gather}
Eliminating  $v_0$ and $v_1$ in terms of $v$ we obtain the
$r$-dDym equation
\begin{gather}\label{r-dDym}
v_{t_2}=\frac{3-r}{(2-r)^2}v^{r-1}\big(v^{2-r}\partial_x^{-1}(v^{r-2}v_{t_1})\big)_{t_1}.
\end{gather}

\section{The $S$-functions for the $r$-dmKP and $r$-dDym hierarchies}

In this section we shall consider the relation \eqref{lax.func} as
a univalent map $p\mapsto L=L(p)$, we shall also use its inverse
$L\mapsto p=p(L)$. We use the notation $\bt:=(t_1,t_2,\dots)$.

\subsection{The $S$-function as a potential}

We introduce potential functions, $S(L,x,\bt)$ and $\tilde
S(\tilde L,x,\bt)$, for
\[
\omega_n(L,x,\bt):=\Omega_n(p(L,x,\bt),x,\bt) \text{ and }
\tilde\omega_n(L,x,\bt):=\tilde \Omega_n(p(\tilde L,x,\bt),x,\bt).
\]
 First, we show that
\begin{pro}
The following identities
  \begin{align*}
    \frac{\partial\omega_n}{\partial t_m}&=\frac{\partial\omega_m}{\partial
    t_n},& \frac{\partial\omega_n}{\partial x}&=p^{-r}\frac{\partial p}{\partial
    t_n}\\
     \frac{\partial\tilde\omega_n}{\partial t_m}&=\frac{\partial\tilde \omega_m}{\partial
    t_n},& \frac{\partial\tilde \omega_n}{\partial x}&=p^{-r}\frac{\partial p}{\partial
    t_n}
  \end{align*}
  hold.
\end{pro}
\begin{proof}
 Let us compute the $t_m$-derivative of $\omega_n$:
  \[
\frac{\partial\omega_n}{\partial
t_m}=\frac{\partial\Omega_n(p(L,x,\bt),x,\bt)}{\partial t_m}=
\frac{\partial\Omega_n}{\partial
p}(p(L,x,\bt),x,\bt)\frac{\partial p}{\partial
t_m}(L,x,\bt)+\frac{\partial\Omega_n}{\partial
t_m}(p(L,x,\bt),x,\bt)
\]
and of
\[
p=p(L(p,x,\bt),x,\bt)
\]
to get
\[
\frac{\partial p}{\partial t_m}=-\frac{\partial p}{\partial
L}\frac{\partial L}{\partial t_m}=-\frac{\partial p}{\partial
L}\{\Omega_m,L\}=-p^r\frac{\partial p}{\partial
L}\Big(\frac{\partial \Omega_m}{\partial p }\frac{\partial
L}{\partial x }-\frac{\partial \Omega_m}{\partial x}\frac{\partial
L}{\partial p }\Big)= p^r\Big(\frac{\partial \Omega_m}{\partial p
}\frac{\partial p}{\partial x }+\frac{\partial \Omega_m}{\partial
x}\Big).
\]
Thus, we deduce
\[
\frac{\partial\omega_n}{\partial t_m}=p^r\Big(\frac{\partial
\Omega_n}{\partial p }\frac{\partial \Omega_m}{\partial p
}\frac{\partial p}{\partial x }+\frac{\partial \Omega_n}{\partial
p }\frac{\partial \Omega_m}{\partial x}\Big)+\frac{\partial
\Omega_n}{\partial t_m }
\]
and
\[
\frac{\partial\omega_n}{\partial
t_m}-\frac{\partial\omega_m}{\partial
t_n}=\{\Omega_n,\Omega_m\}+\frac{\partial \Omega_n}{\partial t_m
}-\frac{\partial\Omega_m}{\partial t_n}=0
\]
as $\Omega_n$ has zero curvature. Observe that
\[
p^{-r}\frac{\partial p}{\partial t_n}=\frac{\partial
\Omega_n}{\partial p }\frac{\partial p}{\partial x
}+\frac{\partial \Omega_n}{\partial
x}=\frac{\partial\omega_n}{\partial x}.
\]
The proof for the remaining cases is performed as above.

\end{proof}
Therefore, we have proven the local existence of functions
$S(L,x,\bt)$ and $\tilde S(\tilde L,x,\bt)$ such that
\begin{gather}\label{S_def}
\frac{\partial S}{\partial t_n}=\omega_n,\quad \frac{\partial
S}{\partial x}=\Pi_r,\quad \frac{\partial \tilde S}{\partial
t_n}=\tilde\omega_n,\quad \frac{\partial \tilde S}{\partial
x}=\Pi_r
\end{gather}
where
\begin{gather}\label{Pir}
\Pi_r:=\begin{cases}
 \dfrac{p^{1-r}}{1-r}, & r\neq 1,\\
 \log p,& r=1.
\end{cases}
 \end{gather}
Notice that
\[
p^r\frac{\d\Pi_r}{\d p}=1.
\]

 We refer to functions satisfying the above
equations \eqref{S_def} as $S$-functions.

\subsection{The $S$-functions and its connection with the Orlov functions}
The so called Orlov functions $M$ and $\tilde M$  are
characterized by the following properties
\begin{enumerate}
  \item They have an expansion of the form
\begin{gather}\label{Orlov.funcs}
\begin{aligned}
M=&\cdots+ (3-r)t_2L^2+(2-r)t_1L+x+w_1(x)L^{-1}+w_2(x)L^{-2}+\cdots,& L&\to \infty,\\
\tilde M=&\cdots+(3-r) t_2 \tilde L^{2}+(2- r)  t_1 \tilde
L+\tilde w_0(x)+\tilde w_1(x) \tilde L^{-1}+ \tilde w_2(x) \tilde
L^{-2}+\cdots, & \tilde L&\to \infty.
\end{aligned}
\end{gather}
Observe that when $r=1$ then $\tilde w_0=x$. \item They are
canonically conjugated to $L$ and $\tilde L$, respectively; i.e.,
\begin{gather}\label{commutation}
\{L,M\}=L^r \text{  and } \{\tilde L,\tilde M\}=\tilde L^r.
\end{gather}
\item Satisfy the Lax equations
\begin{gather}\label{Lax.eqs.M}
\begin{aligned}
\frac{\partial M}{\partial t_n}&=\{\Omega_n,M\},&
\frac{\partial\tilde M}{\partial t_n}&=\{\tilde\Omega_n,\tilde
M\}.
\end{aligned}
\end{gather}
\end{enumerate}

For $r\neq 1$ weshall show that the following functions
\begin{align}
\label{S}S(L,x,\bt)&=\cdots+t_2L^{3-r}+t_1L^{2-r}+\frac{x}{1-r}L^{1-r}+\sum_{n=1}^\infty
S_n(x,\bt)L^{-n+1-r},\quad S_n:=\frac{1}{-n+1-r}w_n(x,\bt)\\
\label{barS}\tilde S(\tilde L,x,\bt)&= \cdots+t_2\tilde
L^{3-r}+t_1\tilde L^{2-r}+\sum_{n=0}^\infty \tilde
S_n(x,\bt)\tilde L^{-n+1-r},\quad \tilde
S_n:=\frac{1}{-n+1-r}\tilde w_n(x,\bt)
\end{align}
are  $S$-functions. For $r=1$ these functions are
\begin{align*}
S(L,x,\bt)&=\cdots+t_2L^{2}+t_1L+x\log L+\sum_{n=1}^\infty
S_n(x,\bt)L^{-n},\quad S_n:=-\frac{1}{n}w_n(x,\bt),\\
\label{barS}\tilde S(\tilde L,x,\bt)&= \cdots+t_2\tilde
L^{2}+t_1\tilde L+x\log \tilde L+\sum_{n=1}^\infty \tilde
S_n(x,\bt) \tilde L^{-n},\quad \tilde S_n:=-\frac{1}{n}\tilde
w_n(x,\bt).
\end{align*}
The role of $S$ and $\tilde S$  as a generating functions is
encoded in the following formulae
\[
M=L^r\frac{\partial S}{\partial L},\quad \tilde M=\tilde
L^r\frac{\partial \tilde S}{\partial \tilde L}.
\]

\begin{pro}\label{Stn}
The functions $S(L,x,\bt)$ and $\tilde S(\tilde L,x,\bt)$ given by
\eqref{S} and \eqref{barS} are $S$-functions; i.e.,
  \begin{gather}\label{Seq}
  \begin{aligned}
   \frac{\partial S}{\partial x}&=\Pi_r,&
    \frac{\partial \tilde S}{\partial x}&=\Pi_r,\\
    \frac{\partial S}{\partial t_n}&=\omega_n,& \frac{\partial \tilde S}{\partial t_n}&=\tilde
    \omega_n, & n&=1,2,\dots.
  \end{aligned}
  \end{gather}
\end{pro}

\begin{proof}

Let us prove that
\[
\frac{\partial S}{\partial x}=\Pi_r.
\]
Observe that according to \eqref{Orlov.funcs}
\[
\frac{\partial M}{\partial x}= \frac{\partial M}{\partial
L}\frac{\partial L}{\partial x}+1+\sum_{n=1}^\infty \frac{\partial
w_n}{\partial x}L^{-n}.
\]
 Thus, assuming that $p,
L\in\C$ and taking a small circle $\gamma$ centered at $L=0$ in
the complex $L$-plane we have,
\begin{align*}
\frac{\partial w_m}{\partial x}=&\int_\gamma\frac{\d L}{2\pi
\text{i}}\Big(L^{m-1}\frac{\partial M}{\partial
x}-L^{m-1}\frac{\partial M}{\partial L}\frac{\partial
L}{\partial x}\Big),& \\
=&\int_\gamma\frac{\d p}{2\pi
\text{i}}L^{m-1}(M_xL_p-L_xM_p)&\text{change of variables
$L=L(p)$}
\\&&\text{and $M_LL_p=M_p$}\\ =&\int_\gamma\frac{\d p}{2\pi
\text{i}}L^{m-1+r}p^{-r}& \text{in virtue of \eqref{commutation}}.
\end{align*}
For $r\neq 1$ we have
\begin{align*}
\frac{\partial w_m}{\partial x}=&(-m+1-r)\int_\gamma\frac{\d
p}{2\pi
\text{i}}L^{m-1+r-1}L_p\frac{p^{1-r}}{1-r}&\text{integration by parts}\\
=&(-m+1-r)\int_\gamma\frac{\d L}{2\pi
\text{i}}L^{m-1+r-1}\frac{p(L)^{1-r}}{1-r}&\text{change of
variables $p=p(L)$}.
\end{align*}
 In the space $\mathcal L$  of Laurent series in $L$ we have a resolution of the identity
$1_{\mathcal L}=\varpi_\geq+\varpi_<$, associated with the
splitting in powers of $L$ of greater or equal degree than $1-r$,
say $\mathcal{L}_{\geq}$, and or less order, $\mathcal{L}_<$

\[
\varpi_\geq(p^{1-r})=L^{1-r},\quad \varpi_<
f=\sum_{m<1-r}\Big(\int_\gamma\frac{\d L}{2\pi
\text{i}}L^{-m-1}f(L)\Big)L^m,\quad\forall f\in\mathcal{L}
\] we get
\begin{align*}
  \frac{\partial S(L,x,\bt)}{\partial x}&=\frac{L^{1-r}}{1-r}+\sum_{n=1}^\infty
\frac{1}{-n+1-r}w_{n,x}(x,\bt)L^{-n+1-r}\\
&=\frac{L^{1-r}}{1-r}+\sum_{m<1-r}\Big(\int_\gamma\frac{\d
L}{2\pi \text{i}}L^{-m-1}\frac{p(L)^{1-r}}{1-r}\Big)L^m\\
&=\varpi_\geq\Big(\frac{p^{1-r}}{1-r}\Big)+\varpi_<\Big(\frac{p^{1-r}}{1-r}\Big)=\frac{p^{1-r}}{1-r}.
\end{align*}
For $r=1$ we have
\[
\frac{\partial w_m}{\partial x}=\int_\gamma\frac{\d \log p}{2\pi
\text{i}}L^{m},
\]
that together with the assumption, suggested by \eqref{lax.func},
\[
\log p=\log L-\Lambda, \quad \Lambda=u_0
L^{-1}+\big(u_1+\frac{u_0^2}{2}\big)L^{-2}+\cdots
\]
allows us to deduce the relations
\[
\frac{\partial w_m}{\partial x}=\int_\gamma\frac{\d L}{2\pi
\text{i}}L^{m-1}-\int_\gamma\frac{\d L}{2\pi
\text{i}}\Lambda_LL^{m}.
\]
Notice that the first term in the r.h.s cancels as $m\geq 1$, then
an integration by parts of the second term in the r.h.s leads to
\[
\frac{\partial w_m}{\partial x}=m\int_\gamma\frac{\d L}{2\pi
\text{i}}\Lambda(L) L^{m-1}.
\]
Hence, we compute
\begin{align*}
  \frac{\partial S(L,x,\bt)}{\partial x}&=\log L-\sum_{m=1}^\infty
\frac{1}{m}w_{m,x}(x,\bt)L^{-m}\\
&=\log L-\sum_{m<0}\Big(\int_\gamma\frac{\d
L}{2\pi \text{i}}L^{-m-1}\Lambda(L)\Big)L^m\\
&=\log L-\Lambda=\log p.
\end{align*}

  Let us  prove the relation $\dfrac{\partial S}{\partial
  t_n}=\omega_n$. Observe that, according to \eqref{S} we have
\[
\frac{\partial S}{\partial t_n}=L^{n+1-r}+\sum_{m=1}^\infty
\frac{1}{-m+1-r}\frac{\partial w_m}{\partial t_n}L^{-m+1-r},
\]
now, from \eqref{Orlov.funcs} we deduce
\begin{gather}\label{MM}
\frac{\partial M}{\partial t_n}=\frac{\partial M}{\partial
L}\frac{\partial L}{\partial t_n}+(n+1-r)L^{n}+\sum_{m=1}^\infty
\frac{\partial w_m}{\partial t_n}L^{-m}.\end{gather}

Thus
\begin{align*}
\frac{\partial w_m}{\partial t_n}=&\int_\gamma\frac{\d L}{2\pi
\text{i}}\Big(L^{m-1}\frac{\partial M}{\partial
t_n}-L^{m-1}\frac{\partial M}{\partial L}\frac{\partial
L}{\partial t_n}\Big),& \text{from \eqref{MM}}\\
=&\int_\gamma\frac{\d L}{2\pi
\text{i}}\Big(L^{m-1}\{\Omega_n,M\}-L^{m-1}\frac{\partial
M}{\partial L}\{\Omega_n,L\}\Big), & \text{derived from \eqref{Lax.eqs} and \eqref{Lax.eqs.M}}\\
=&\int_\gamma\frac{\d L}{2\pi
\text{i}}\Big(L^{m-1}p^r(\Omega_{n,p}M_x-\Omega_{n,x}M_p)\\&\quad\quad\quad
-L^{m-1}\frac{\partial
M}{\partial
L}p^r(\Omega_{n,p}L_x-\Omega_{n,x}L_p)\Big)& \text{see \eqref{bracket}}\\
=&\int_\gamma\frac{\d p}{2\pi
\text{i}}L^{m-1}p^r\Omega_{n,p}(M_xL_p-L_xM_p)&\text{change of
variables $L=L(p)$}
\\&&\text{and $M_LL_p=M_p$}\\ =&\int_\gamma\frac{\d p}{2\pi
\text{i}}L^{m-1+r}\Omega_{n,p}& \text{in virtue of \eqref{commutation}}\\
=&(-m+1-r)\int_\gamma\frac{\d p}{2\pi
\text{i}}L^{m-1+r-1}L_p\Omega_n&\text{integration by parts}\\
=&(-m+1-r)\int_\gamma\frac{\d L}{2\pi
\text{i}}L^{m-1+r-1}\omega_n&\text{change of variables $p=p(L)$}
\end{align*}
and therefore
\[
\frac{\partial S}{\partial t_n}=L^{n+1-r}+\sum_{m<1-r}
\Big(\int_\gamma\frac{\d L}{2\pi
\text{i}}L^{-m-1}\omega_n\Big)L^{m}.
\]
Now,
\[
\int_\gamma\frac{\d L}{2\pi
\text{i}}L^{-m-1}\omega_n=\int_\gamma\frac{\d L}{2\pi
\text{i}}L^{-m-1}(L^{n+1-r}-P_<L^{n+1-r})=- \int_\gamma\frac{\d
L}{2\pi \text{i}}L^{-m-1}P_<L^{n+1-r}
\]
as $m\neq n+1-r$ for $m<1-r$. So that
\begin{gather*}
\frac{\partial S}{\partial
t_n}=\omega_n+P_<(L^{n+1-r})-\sum_{m<1-r} \Big(\int_\gamma\frac{\d
L}{2\pi \text{i}}L^{-m-1}P_<(L^{n+1-r})\Big)L^{m},
\end{gather*}
but
\[
P_<(L^{n+1-r})=\sum_{m\in\Z} \Big(\int_\gamma\frac{\d L}{2\pi
\text{i}}L^{-m-1}P_<(L^{n+1-r})\Big)L^{m}=\sum_{m<1-r}
\Big(\int_\gamma\frac{\d L}{2\pi
\text{i}}L^{-m-1}P_<(L^{n+1-r})\Big)L^{m}
\]
and the result follows. For $r\neq 1$, let us prove that $
\dfrac{\partial \tilde S}{\partial x}=\dfrac{p^{1-r}}{1-r}$;
equations \eqref{Orlov.funcs} imply
\[
\frac{\partial \tilde  M}{\partial x}= \frac{\partial \tilde
M}{\partial \tilde L}\frac{\partial \tilde L}{\partial
x}+\sum_{n=0}^\infty \frac{\partial \tilde w_n}{\partial x}\tilde
L^{-n}.
\]
and hence
\begin{align*}
\frac{\partial \tilde w_m}{\partial x}=&\int_\gamma\frac{\d
L}{2\pi \text{i}}\Big(\tilde L^{m-1}\frac{\partial \tilde
M}{\partial x}-\tilde L^{m-1}\frac{\partial \tilde M}{\partial
\tilde L}\frac{\partial
\tilde L}{\partial x}\Big),& \\
=&\int_\gamma\frac{\d p}{2\pi \text{i}}\tilde L^{m-1}(\tilde
M_x\tilde L_p-\tilde L_x\tilde M_p)&\text{change of variables
$\tilde L=\tilde L(p)$}
\\&&\text{and $\tilde M_{\tilde L}\tilde L_p=\tilde M_p$}\\ =&\int_\gamma\frac{\d p}{2\pi
\text{i}}\tilde L^{m-1+r}p^{-r}& \text{in virtue of \eqref{commutation}}\\
=&(-m+1-r)\int_\gamma\frac{\d p}{2\pi
\text{i}}\tilde L^{m-1+r-1}\tilde L_p\frac{p^{1-r}}{1-r}&\text{integration by parts}\\
=&(-m+1-r)\int_\gamma\frac{\d \tilde L}{2\pi \text{i}}\tilde
L^{m-1+r-1}\frac{p^{1-r}}{1-r}&\text{change of variables
$p=p(\tilde L)$}.
\end{align*}
We now proceed computing
\begin{align*}
  \frac{\partial \tilde S(\tilde L,x,\bt)}{\partial x}=\sum_{n=0}^\infty
\frac{1}{-n+1-r}\tilde w_{n,x}(x,\bt)\tilde L^{-n+1-r}
=\sum_{m\leq 1-r}\Big(\int_\gamma\frac{\d \tilde L}{2\pi
\text{i}}\tilde L^{-m-1}\frac{p^{1-r}}{1-r}\Big)\tilde L^m,
\end{align*}
noting that $ \dfrac{p^{1-r}}{1-r}\in\g_\leq$ we get the desired
result. For $r=1$ we have
\[
\frac{\partial \tilde w_m}{\partial x}=\int_\gamma\frac{\d \log
p}{2\pi \text{i}}\tilde L^{m}
\]
assuming, as suggested by \eqref{llax.func}, that
\[
\log p=\log \tilde L-\tilde \Lambda,\quad \tilde \Lambda:=\log
v+\frac{v_0}{v}\tilde
L^{-1}+\Big(v_1+\frac{v_0^2}{2v^2}\Big)\tilde L^{-2}+\cdots
\]
we have
\begin{align*}
\frac{\partial \tilde w_m}{\partial x}&=\int_\gamma\frac{\d \tilde
L}{2\pi \text{i}}\tilde L^{m-1}-\int_\gamma\frac{\d  \tilde
L}{2\pi \text{i}}\tilde \Lambda_{\tilde L}\tilde L^m\\
&=m\int_\gamma\frac{\d  \tilde L}{2\pi \text{i}}\tilde \Lambda
L^{m-1},\quad m\geq 1.
\end{align*}
Thus, we calculate
\begin{align*}
  \frac{\partial \tilde S(\tilde L,x,\bt)}{\partial x}&=\log\tilde L-\sum_{m=1}^\infty
\frac{1}{m}\tilde w_{m,x}(x,\bt)\tilde L^{-m} =\log\tilde
L-\sum_{m<0}\Big(\int_\gamma\frac{\d \tilde L}{2\pi
\text{i}}\tilde L^{-m-1}\tilde\Lambda(\tilde L)\Big)\tilde
L^m\\&=\log\tilde L-\tilde \Lambda=\log p.
\end{align*}


To check the formula $\dfrac{\partial \tilde S}{\partial
t_n}=\tilde \omega_n$, we procede in a similar way. Equation
\eqref{barS} implies that
\[
\frac{\partial \tilde S}{\partial t_n}=\tilde
L^{n+1-r}+\sum_{m=0}^\infty \frac{1}{-m+1-r}\frac{\partial \tilde
w_m}{\partial t_n}\tilde L^{-m+1-r},
\]
now, from \eqref{Orlov.funcs} we deduce
\begin{gather}\label{tildeMM}
\frac{\partial \tilde M}{\partial t_n}=\frac{\partial \tilde
M}{\partial \tilde L}\frac{\partial \tilde L}{\partial
t_n}+(n+1-r)\tilde L^{n}+\sum_{m=0}^\infty \frac{\partial \tilde
w_m}{\partial t_n}\tilde L^{-m}.\end{gather}

 As for the $r$-dmKP case we have the following chain of
 observations
\begin{align*}
\frac{\partial \tilde w_m}{\partial t_n}=&\int_\gamma\frac{\d
\tilde L}{2\pi \text{i}}\Big(\tilde L^{m-1}\frac{\partial \tilde
M}{\partial t_n}-\tilde L^{m-1}\frac{\partial \tilde M}{\partial
\tilde L}\frac{\partial
\tilde L}{\partial t_n}\Big),& \text{from \eqref{tildeMM}}\\
=&\int_\gamma\frac{\d \tilde L}{2\pi \text{i}}\Big(\tilde
L^{m-1}\{\tilde \Omega_n,\tilde M\}-\tilde L^{m-1}\frac{\partial
\tilde M}{\partial \tilde L}\{\tilde \Omega_n,\tilde L\}\Big), & \text{derived from \eqref{Lax.eqs} and \eqref{Lax.eqs.M}}\\
=&\int_\gamma\frac{\d p}{2\pi \text{i}}\tilde
L^{m-1}p^r\tilde\Omega_{n,p}(\tilde M_x\tilde L_p-\tilde L_x\tilde
M_p)&\text{change of variables $\tilde L=\tilde L(p)$}
\\
=&(-m+1-r)\int_\gamma\frac{\d p}{2\pi
\text{i}}\tilde L^{m-1+r-1}\tilde L_p\tilde\Omega_n&\text{integration by parts}\\
=&(-m+1-r)\int_\gamma\frac{\d \tilde L}{2\pi \text{i}}\tilde
L^{m-1+r-1}\tilde\Omega_n&\text{change of variables $p=p(\tilde
L)$}
\end{align*}
that lead to
\[
\frac{\partial \tilde S}{\partial t_n}=\tilde
L^{n+1-r}+\sum_{m\leq 1-r} \Big(\int_\gamma\frac{\d \tilde L}{2\pi
\text{i}}\tilde L^{-m-1}\tilde \Omega_n\Big)\tilde L^{m}.
\]
Also we have
\[
\int_\gamma\frac{\d \tilde L}{2\pi \text{i}}\tilde L^{-m-1}\tilde
\Omega_n=\int_\gamma\frac{\d \tilde L}{2\pi \text{i}}\tilde
L^{-m-1}(\tilde L^{n+1-r}-P_\leq\tilde L^{n+1-r})=-
\int_\gamma\frac{\d \tilde L}{2\pi \text{i}}L^{-m-1}P_\leq\tilde
L^{n+1-r}
\]
as $m\neq n+1-r$ for $m\leq 1-r$. Therefore,
\begin{gather*}
\frac{\partial \tilde S}{\partial t_n}=\tilde
\Omega_n+P_\leq(\tilde L^{n+1-r})-\sum_{m\leq 1-r}
\Big(\int_\gamma\frac{\d \tilde L}{2\pi \text{i}}\tilde
L^{-m-1}P_\leq(\tilde L^{n+1-r})\Big)\tilde L^{m},
\end{gather*}
and from
\[
P_\leq(\tilde L^{n+1-r})=\sum_{m\in\Z} \Big(\int_\gamma\frac{\d
\tilde L}{2\pi \text{i}}\tilde L^{-m-1}P_\leq(\tilde
L^{n+1-r})\Big)\tilde L^{m}=\sum_{m\leq 1-r}
\Big(\int_\gamma\frac{\d\tilde L}{2\pi \text{i}}\tilde
L^{-m-1}P_\leq(\tilde L^{n+1-r})\Big)\tilde L^{m}
\]
we arrive to the claimed result.
\end{proof}

\subsection{The $S$-function formulation of the integrable hierarchies}
Here we show that the integrable hierarchies can be formulated in
terms of $S$-functions. This is a key observation for the
reduction procedure that we shall present in the next section.

\begin{pro}
  Let $L$ and $\tilde L$ be functions with expansions as given in
  \eqref{lax.func} and \eqref{llax.func}, respectively, and $\Pi_r$ as defined in \eqref{Pir}. Let $S(L,x,\bt)$
  and $\tilde S(\tilde L,x,\bt)$ be $S$-functions; i.e.
\begin{gather*}
  \begin{aligned}
   \frac{\partial S}{\partial x}&=\Pi_r,&
    \frac{\partial \tilde S}{\partial x}&=\Pi_r,\\
    \frac{\partial S}{\partial t_n}&=P_\geq(L^{n+1-r}),& \frac{\partial \tilde S}{\partial t_n}&=
    P_>(\tilde L^{n+1-r}), & n&=1,2,\dots.
  \end{aligned}
  \end{gather*}
Then, $L$ and $\tilde L$ do satisfy the $r$-dmKP and $r$-dDym
hierarchies \eqref{Lax.eqs} and \eqref{LLax.eqs}, respectively;
i.e.,
\[
\frac{\partial L}{\partial t_n}=\{P_\geq(L^{n+1-r}),L\}, \quad
\frac{\partial \tilde L}{\partial t_n}=\{P_>(\tilde
L^{n+1-r}),\tilde L\}.
\]

  \end{pro}
\begin{proof}
From $S_x=\Pi_r$ we deduce
  \[
\frac{\partial p}{\partial t_n}=p^r \frac{\partial^2 S}{\partial
t_n\partial
x}=p^r(\Omega_n(p(L,x,\bt),x,\bt))_x=p^r(\Omega_{n,p}p_x+\Omega_x).
  \]
  Thus,
  \[
\frac{\partial L}{\partial t_n}=-\frac{\partial L}{\partial
p}\frac{\partial p}{\partial
t_n}=-L_pp^r(\Omega_{n,p}p_x+\Omega_x)=
p^r(-\Omega_{n,p}L_x+\Omega_xL_p)=\{\Omega_n,L\},
  \]
  as claimed. The statement for the $r$-dDym hierarchy
  follows analogously.
\end{proof}

%
%


\section{Reductions}

The reductions we consider here are motivated by the reductions we
studied in \cite{otros}  of the dispersionless KP hierarchy. We
assume that the $\bt$-dependence appears always in terms of
$\bu=(U_1,\dots,U_N)$, a set of $N$ functions of $\bt$ and $x$.
This dependence is defined trough the following equations for the
function $p=p(L,\bu)$ or $p=p(\tilde L,\bu)$
\begin{gather}\label{reduction.p}
\frac{\partial p}{\partial U_i}=R_i(p,\bu),\quad i=1,\dots,N,
\end{gather}
which in terms of the Lax functions are
\begin{align}\label{reduction.L}
\frac{\partial L}{\partial U_i}+R_i(p,\bu)\frac{\partial L}{\partial p}&=0,\quad i=1,\dots,N,\\
\label{reduction.tilde.L} \frac{\partial \tilde L}{\partial
U_i}+\tilde R_i(p,\bu)\frac{\partial \tilde L}{\partial
p}&=0,\quad i=1,\dots,N.
\end{align}

 We shall assume that the compatibility conditions for
\eqref{reduction.p} are fulfilled; i.e.  both sets of functions
$\{R_i\}_{i=1}^N$ and $\{\tilde R_i\}_{i=1}^N$ fulfill
\begin{gather}\label{compatibility.R}
\frac{\partial R_i}{\partial U_j}-\frac{\partial R_j}{\partial
U_i}+R_j\frac{\partial R_i}{\partial p}-R_i\frac{\partial
R_j}{\partial p }=0.
\end{gather}

 We will also suppose that $R_i$ and $\tilde R_i$, $i=1,\dots, N$, are rational functions with
$N$ simple poles, $\pi_i=\pi_i(\bu)$ and
$\tilde\pi_i=\tilde\pi_i(\bu)$ $i=1,\dots,N$, respectively.
Recalling the expansions \eqref{lax.func} and \eqref{llax.func}
and taking into account formulae \eqref{reduction.L} and
\eqref{reduction.tilde.L}, we request to $R_i$ to be of order
$O(1)$ when $p\to\infty$  and $\tilde R_i$ of order $O(p)$ when
$p\to\infty$. Hence, our functions $R$ will be of the form
\begin{align}\label{R.form.0}
   R_i(\bu)&=
p\sum_{j=1}^N\dfrac{\rho_{ij}(\bu)}{p-\pi_j(\bu)},\\
\label{tilde.R.form.0} \tilde
R_i(\bu)&=p^2\sum_{j=1}^N\dfrac{\tilde\rho_{ij}(\bu)}{p-\tilde\pi_j(\bu)}.
\end{align}

The asymptotic behaviors for $p\to\infty$ are
\[
\begin{aligned}
  R_i&=R_{i,0}+R_{i,1}p^{-1}+R_{i,2}p^{-2}+\cdots,\\
  \tilde R_i&=\tilde R_{i,0}p+\tilde R_{i,1}+\tilde R_{i,2}p^{-1}+\cdots,
\end{aligned}
\]
 were we have used the following notation
\[
R_{i,n}:=\sum_{j=1}^N\rho_{ij}\pi_j^n,\quad \tilde
R_{i,n}:=\sum_{j=1}^N\tilde\rho_{ij}\tilde\pi_j^n.
\]


The equation \eqref{reduction.L} together with \eqref{R.form.0},
in the $r$-dmKP hierarchy case, imply
\begin{gather}\label{relations.KP}
  \begin{aligned}
\frac{\partial u_0}{\partial U_i}&=-R_{i,0},\\
\frac{\partial u_1}{\partial U_i}&=-R_{i,1},\\
\frac{\partial u_2}{\partial U_i}&=-R_{i,2}+R_{i,0}u_1,\\
&\vdots
  \end{aligned}
  \end{gather}
Observe that all the coefficients $u_n$ are expressed recursively
in terms of the functions $\{\pi_k,\rho_{ik}\}$ defining $R_i$.
The equation \eqref{reduction.tilde.L} together with
\eqref{tilde.R.form.0}, in the $r$-dDym hierarchy case, leads to
\begin{gather}\label{relations.dym}
  \begin{aligned}
\frac{\partial v}{\partial U_i}&=-\tilde R_{i,0}v,\\
\frac{\partial v_0}{\partial U_i}&=-\tilde R_{i,1}v,\\
\frac{\partial v_1}{\partial U_i}&=-\tilde R_{i,2}v+\tilde R_{i,0}v_1,\\
&\vdots
  \end{aligned}
  \end{gather}

Observe that all the coefficients $v$ and $v_n$ are expressed
recursively in terms of the functions
$\tilde\pi_k,\tilde\rho_{ik}\}$ defining $R_i$.

\subsection{On the compatibility conditions}

We now discuss the compatibility conditions for
\eqref{compatibility.R}.
\paragraph{$r$-dDym}
The compatibility equations for \eqref{compatibility.R} with the
choice \eqref{tilde.R.form.0} are
\begin{subequations}
 \begin{align}
 \label{compatibility.dym.0.2}
  \tilde\rho_{il}\frac{\partial \tilde\pi_l}{\partial U_j}-
\tilde \rho_{jl}\frac{\partial \tilde\pi_l}{\partial U_i}
  &=\sum_{k\neq l}\frac{\tilde\rho_{ik}\tilde\rho_{jl}-\tilde\rho_{il}\tilde\rho_{jk}}{\tilde\pi_k-\tilde\pi_l}\tilde\pi_l^2,\\
  \label{compatibility.dym.0.3}
\frac{\partial \tilde\rho_{il}}{\partial U_j}-\frac{\partial
\tilde\rho_{jl}}{\partial U_i}
  &=2\sum_{k\neq
  l}\frac{\tilde\rho_{ik}\tilde\rho_{jl}-\tilde\rho_{il}\tilde\rho_{jk}}{(\tilde\pi_l-\tilde\pi_k)^2}\tilde\pi_k\tilde\pi_l.
\end{align}
\end{subequations}
 From equations \eqref{compatibility.dym.0.2}
 and \eqref{compatibility.dym.0.3} we deduce
\begin{pro}\label{potential.dym}
  There exist a pair of potentials $\tilde\sigma$ and $\tilde\rho$
  such that
  \begin{align}
\label{potential1.dym} \tilde R_{i,0}&=-\frac{\partial
\tilde\sigma}{\partial U_i},\\
\label{potential2.dym}\tilde R_{i,1}&=-\frac{\partial\tilde
\rho}{\partial U_i}-\tilde\rho\frac{\partial
\tilde\sigma}{\partial U_i}.
  \end{align}
\end{pro}
\begin{proof}
  Firstly, we observe that
  \[
  \frac{\partial R_{i,0}}{\partial U_j}-\frac{\partial R_{j,0}}{\partial
  U_i}=\sum_{l=1}^N\Big(\frac{\partial\tilde\rho_{il}}{\partial U_j}-\frac{\partial\tilde\rho_{jl}}{\partial U_i}\Big)
  =2\sum_{\substack{l=1,\dots,N\\k\neq
  l}}\frac{\tilde\rho_{ik}\tilde\rho_{jl}-\tilde\rho_{il}\tilde\rho_{jk}}{(\tilde\pi_l-\tilde\pi_k)^2}\tilde\pi_k\tilde\pi_l=0
  \]
  Secondly, we evaluate
\begin{align*}
  \frac{\partial R_{i,1}}{\partial U_j}-R_{j,0}R_{i,1}-
\Big(\frac{\partial R_{j,1}}{\partial U_i}-R_{i,0}R_{j,1}\Big)&=
\sum_{\substack{l=1,\dots,N\\k\neq
  l}}\Big[(
\tilde\rho_{ik}\tilde\rho_{jl}-\tilde\rho_{il}\tilde\rho_{jk})
\Big(
\frac{2\tilde\pi_l^2\tilde\pi_k}{(\tilde\pi_l-\tilde\pi_k)^2}
-\frac{\tilde\pi_l^2}{\tilde\pi_l-\tilde\pi_k}+\tilde\pi_l \Big)
  \Big]\\
  &=\sum_{\substack{l=1,\dots,N\\k\neq
  l}}(
\tilde\rho_{ik}\tilde\rho_{jl}-\tilde\rho_{il}\tilde\rho_{jk})\tilde\pi_k\tilde\pi_l(\tilde\pi_k+\tilde\pi_l)=0.
\end{align*}
The stated result is a direct consequence of these two equations.
\end{proof}

\paragraph{$r$-dmKP}
The compatibility equations \eqref{compatibility.R} for
\eqref{R.form.0}  are
\begin{subequations}
 \begin{align}
 \label{compatibility.rp.0.2}
  \rho_{il}\frac{\partial \pi_l}{\partial U_j}-
  \rho_{jl}\frac{\partial \pi_l}{\partial U_i}
  &=\sum_{k\neq l}\frac{\rho_{ik}\rho_{jl}-\rho_{il}\rho_{jk}}{\pi_k-\pi_l}\pi_l,\\
  \label{compatibility.rp.0.3}
\frac{\partial \rho_{il}}{\partial U_j}-\frac{\partial
\rho_{jl}}{\partial U_i}
  &=\sum_{k\neq
l}\frac{\rho_{ik}\rho_{jl}-\rho_{il}\rho_{jk}}{(\pi_l-\pi_k)^2}(\pi_k+\pi_l).
\end{align}
\end{subequations}
As in Proposition \ref{potential.dym} we can show that
\begin{pro}
  There exist a potential $\rho$
  such that
  \begin{align}
\label{potential1.kp} R_{i,0}&=-\frac{\partial \rho}{\partial
U_i}.
  \end{align}
\end{pro}


The following Proposition shows a connection between the
compatibility conditions for the $r$-dDym and $r$-dmKP equations.
Further, it also shows a connection among these and the
compatibility conditions
\begin{subequations}
 \begin{align}
\label{compatibility.rp.2.1} r_{il}\frac{\partial p_l}{\partial
U_j}- r_{jl}\frac{\partial  p_l}{\partial U_i}
  &=\sum_{k\neq l}\frac{r_{jl}r_{ik}- r_{il} r_{jk}}{p_k- p_l},\\
  \label{compatibility.rp.3.1}
\frac{\partial  r_{il}}{\partial U_j}-\frac{\partial
r_{jl}}{\partial U_i}
  &=2\sum_{k\neq
l}\frac{ r_{jl} r_{ik}- r_{il}r_{jk}}{( p_l-p_k)^2}.
\end{align}
\end{subequations} for similar reductions of the
dispersionless KP hierarchy, that we discussed in some length in
\cite{otros}.
\begin{pro}
  \begin{enumerate}
    \item If $\tilde\pi_i$ and $\tilde\rho_{ij}$ solves the
    compatibility conditions \eqref{compatibility.dym.0.2} and
    \eqref{compatibility.dym.0.3} then
    \begin{gather}\label{dym-to-mkp}
          \pi_i=\Exp{\tilde\sigma}\tilde\pi_i,\quad
    \rho_{ij}=\Exp{\tilde\sigma}\tilde\rho_{ij}\tilde\pi_j
    \end{gather}
   solves the compatibility conditions
    \eqref{compatibility.rp.0.2} and \eqref{compatibility.rp.0.3}.
    Moreover, we may take as the potential $\rho$ the following
    function
    \[
    \rho=\Exp{\tilde\sigma}\tilde\rho.
    \]
    \item If $\pi_i$ and $\rho_{ij}$ solves the
    compatibility conditions   \eqref{compatibility.rp.0.2} and
    \eqref{compatibility.rp.0.3} then
   \begin{gather}\label{mkp-to-kp}
    p_i=\pi_i+\rho,\quad
    r_{ij}=\rho_{ij}\pi_j
    \end{gather}
    solves the compatibility conditions
    \eqref{compatibility.rp.2.1} and \eqref{compatibility.rp.3.1}.
  \end{enumerate}
\end{pro}

\begin{proof}
\begin{enumerate}
  \item
With the expressions \eqref{dym-to-mkp}  and the formulae
\eqref{compatibility.dym.0.2} and \eqref{compatibility.dym.0.3} we
evaluate
\begin{align*}
  \rho_{il}\frac{\partial \pi_l}{\partial U_j}-
  \rho_{jl}\frac{\partial \pi_l}{\partial
  U_i}&=\Exp{2\tilde\sigma}\bigg(\Big(  \tilde\rho_{il}\frac{ \partial \tilde\pi_l}{\partial U_j}-
   \tilde\rho_{jl}\frac{\partial  \tilde\pi_l}{\partial U_i}\Big) \tilde\pi_l-(
   \tilde\rho_{il} \tilde R_{j,0}- \tilde\rho_{jl} \tilde
   R_{i,0})
   \tilde\pi_l^2\bigg)\\
      &=\Exp{2\tilde\sigma}\sum_{k\neq l}(\tilde\rho_{ik}\tilde\rho_{jl}-\tilde\rho_{il}\tilde\rho_{jk})
\tilde\pi_l^2\Big(\frac{\tilde \pi_l}{\tilde \pi_k-\tilde\pi_l}+1
\Big)\\&=\Exp{2\tilde\sigma}\sum_{k\neq
l}(\tilde\rho_{ik}\tilde\rho_{jl}-\tilde\rho_{il}\tilde\rho_{jk})
\tilde \pi_k\tilde \pi_l\frac{\tilde
\pi_l}{\tilde \pi_k-\tilde\pi_l}\\
&=\sum_{k\neq
l}\frac{\rho_{ik}\rho_{jl}-\rho_{il}\rho_{jk}}{\pi_k-\pi_l}\pi_l
\\
\frac{\partial\rho_{il}}{\partial U_j}-
 \frac{\partial  \rho_{jl}}{\partial
  U_i}&=\Exp{\tilde\sigma}\bigg(\frac{\partial\tilde\rho_{il}}{\partial U_j}-
 \frac{\partial  \tilde\rho_{jl}}{\partial
  U_i} -(
   \tilde\rho_{il} \tilde R_{j,0}- \tilde\rho_{jl} \tilde
   R_{i,0})
   \tilde\pi_l +\tilde\rho_{il}\frac{ \partial \tilde\pi_l}{\partial U_j}-
   \tilde\rho_{jl}\frac{\partial  \tilde\pi_l}{\partial U_i}
   \bigg)\\
   &=\Exp{\tilde\sigma}\sum_{k\neq
l}(\tilde\rho_{ik}\tilde\rho_{jl}-\tilde\rho_{il}\tilde\rho_{jk})\Big(
\frac{2\tilde \pi_k\tilde \pi_l^2}{(\tilde
\pi_k-\tilde\pi_l)^2}+\frac{\tilde \pi_k\tilde \pi_l}{\tilde
\pi_k-\tilde\pi_l} \Big)
\\
   &=\Exp{\tilde\sigma}\sum_{k\neq
l}(\tilde\rho_{ik}\tilde\rho_{jl}-\tilde\rho_{il}\tilde\rho_{jk})\tilde
\pi_k\tilde \pi_l \frac{\tilde \pi_k+\tilde \pi_l}{(\tilde
\pi_k-\tilde\pi_l)^2}\\&=\sum_{k\neq
l}\frac{\rho_{ik}\rho_{jl}-\rho_{il}\rho_{jk}}{(\pi_l-\pi_k)^2}(\pi_k+\pi_l)
\end{align*}
and as claimed we have gotten  \eqref{compatibility.rp.0.2} and
\eqref{compatibility.rp.0.3}.

From \eqref{dym-to-mkp} we get
\[
R_{i,0}=\Exp{\tilde\sigma}\tilde R_{i,1};
\]
i.e.,
\[
\frac{\partial\rho}{\partial
U_i}=\Exp{\tilde\sigma}\Big(\frac{\partial\rho}{\partial
U_i}+\tilde\rho\frac{\partial\tilde\sigma}{\partial
U_i}\Big)=\frac{\partial(\Exp{\tilde\sigma}\tilde\rho)}{\partial
U_i}.
\]
    \item We now use \eqref{mkp-to-kp} together with \eqref{compatibility.rp.0.2} and
\eqref{compatibility.rp.0.3} to get
\begin{align*}
  r_{il}\frac{\partial p_l}{\partial U_j}-
  r_{jl}\frac{\partial p_l}{\partial
  U_i}&=\Big(  \rho_{il}\frac{ \partial \pi_l}{\partial U_j}-
   \rho_{jl}\frac{\partial  \pi_l}{\partial U_i}\Big)
   \pi_l-\rho_{il}R_{j,0}+
\rho_{jl}R_{i,0}\\
      &=\sum_{k\neq l}(\rho_{ik}\rho_{jl}-\rho_{il}\rho_{jk})
\pi_l\Big(\frac{ \pi_l}{\pi_k-\pi_l}+1 \Big)\\&=\sum_{k\neq
l}(\rho_{ik}\rho_{jl}-\rho_{il}\rho_{jk})
 \pi_k \pi_l\frac{
1}{ \pi_k-\pi_l}\\
&=\sum_{k\neq l}\frac{r_{ik}r_{jl}-r_{il}r_{jk}}{p_k-p_l}
\\
\frac{\partial r_{il}}{\partial U_j}-
 \frac{\partial  r_{jl}}{\partial
  U_i}&=\Big(\frac{\partial\rho_{il}}{\partial U_j}-
 \frac{\partial  \rho_{jl}}{\partial
  U_i}\Big)\pi_l  +\rho_{il}\frac{ \partial \pi_l}{\partial U_j}-
   \rho_{jl}\frac{\partial  \pi_l}{\partial U_i}
   \\
   &=\sum_{k\neq
l}(\rho_{ik}\rho_{jl}-\rho_{il}\rho_{jk})\Big(
\frac{(\pi_k+\pi_l)\pi_l}{(\pi_k-\pi_l)^2}+\frac{
\pi_l}{\pi_k-\pi_l} \Big)
\\
   &=2\sum_{k\neq
l}(\rho_{ik}\rho_{jl}-\rho_{il}\rho_{jk}) \pi_k\pi_l
\frac{1}{(\pi_k-\pi_l)^2}\\&=2\sum_{k\neq
l}\frac{r_{ik}r_{jl}-r_{il}r_{jk}}{(p_l-p_k)^2}.
\end{align*}

\end{enumerate}
\end{proof}

The inverse statement also holds.

\begin{pro}
\begin{enumerate}
  \item Let $\pi_i,\rho_{ij}$ be solutions of \eqref{compatibility.rp.0.2} and
  \eqref{compatibility.rp.0.3}, then there exists a potential
  function $\tilde\sigma$ such that
  \[
 \frac{\partial \tilde \sigma}{\partial
  U_i}=- \sum_{l=1}^N\frac{\rho_{il}}{\pi_l}.
  \]
  Moreover,
  \[
  \tilde \pi_i=\Exp{-\tilde\sigma}\pi_i,\quad
  \tilde\rho_{ij}=\frac{\rho_{ij}}{\pi_j},
  \]
  provides us with solutions to \eqref{compatibility.dym.0.2} and
  \eqref{compatibility.dym.0.3}.
  \item Let $p_i,r_{ij}$ solutions   \eqref{compatibility.rp.2.1} and \eqref{compatibility.rp.3.1}
   then there exists a potential
  function $\rho$ such that
  \[
  \frac{\partial \rho}{\partial
  U_i}=\sum_{l=1}^N\frac{r_{il}}{\rho-p_l}.
  \]
Moreover,
\[
\pi_i:= p_j-\rho,\quad\rho_{ij}
:=\frac{r_{ij}}{p_j-\rho}
\]
solves the equations \eqref{compatibility.rp.0.2} and
  \eqref{compatibility.rp.0.3}.
\end{enumerate}
\end{pro}

\begin{proof}
  \begin{enumerate}
    \item Let us evaluate
    \begin{align*}
      \frac{\partial}{\partial
      U_j}\Big(\sum_{l=1}^N\frac{\rho_{il}}{\pi_l}\Big)-\frac{\partial}{\partial
      U_i}\Big(\sum_{l=1}^N\frac{\rho_{jl}}{\pi_l}\Big)&=\sum_{l=1}^N
\Big(
\frac{\rho_{il,j}-\rho_{jl,i}}{\pi_l}-\frac{\rho_{il}\pi_{l,j}-\rho_{jl}\pi_{l,i}}{\pi_l^2}\Big)\\
&=\sum_{\substack{l=1,\dots,N\\k\neq
  l}}
\Big((\rho_{ik}\rho_{jl}-\rho_{il}\rho_{jk})\Big(\frac{\pi_k+\pi_l}{\pi_l(\pi_l-\pi_k)^2}-
\frac{\pi_k-\pi_l}{\pi_l(\pi_l-\pi_k)^2}\Big)\Big)\\&=
2\sum_{\substack{l=1,\dots,N\\k\neq
  l}}
\frac{\rho_{ik}\rho_{jl}-\rho_{il}\rho_{jk}}{(\pi_l-\pi_k)^2}=0.
    \end{align*}
    So that, locally, the existence of the mentioned potential
    holds, and the relation \eqref{dym-to-mkp} allows us to identify
    it with $\tilde\sigma$.
Moreover, the identities
\begin{align*}
  \rho_{il}\frac{\partial \pi_l}{\partial U_j}-
  \rho_{jl}\frac{\partial \pi_l}{\partial
  U_i}&=\Exp{2\tilde\sigma}\bigg(\Big(  \tilde\rho_{il}\frac{ \partial \tilde\pi_l}{\partial U_j}-
   \tilde\rho_{jl}\frac{\partial  \tilde\pi_l}{\partial U_i}\Big) \tilde\pi_l-(
   \tilde\rho_{il} \tilde R_{j,0}- \tilde\rho_{jl} \tilde
   R_{i,0})
   \tilde\pi_l^2\bigg)\\\sum_{k\neq
l}\frac{\rho_{ik}\rho_{jl}-\rho_{il}\rho_{jk}}{\pi_k-\pi_l}\pi_l&=\Exp{2\tilde\sigma}\sum_{k\neq
l}(\tilde\rho_{ik}\tilde\rho_{jl}-\tilde\rho_{il}\tilde\rho_{jk})
\tilde \pi_k\tilde \pi_l\frac{\tilde
\pi_l}{\tilde \pi_k-\tilde\pi_l}\\
\frac{\partial\rho_{il}}{\partial U_j}-
 \frac{\partial  \rho_{jl}}{\partial
  U_i}&=\Exp{\tilde\sigma}\bigg(\frac{\partial\tilde\rho_{il}}{\partial U_j}-
 \frac{\partial  \tilde\rho_{jl}}{\partial
  U_i} -(
   \tilde\rho_{il} \tilde R_{j,0}- \tilde\rho_{jl} \tilde
   R_{i,0})
   \tilde\pi_l +\tilde\rho_{il}\frac{ \partial \tilde\pi_l}{\partial U_j}-
   \tilde\rho_{jl}\frac{\partial  \tilde\pi_l}{\partial U_i}
   \bigg)\\
  \sum_{k\neq
l}\frac{\rho_{ik}\rho_{jl}-\rho_{il}\rho_{jk}}{(\pi_l-\pi_k)^2}(\pi_k+\pi_l)
&=\Exp{\tilde\sigma}\sum_{k\neq
l}(\tilde\rho_{ik}\tilde\rho_{jl}-\tilde\rho_{il}\tilde\rho_{jk})\tilde
\pi_k\tilde \pi_l \frac{\tilde \pi_k+\tilde \pi_l}{(\tilde
\pi_k-\tilde\pi_l)^2},
\end{align*}
imply our statements.
    \item The compatibility conditions for
    \[
  \frac{\partial \rho}{\partial
  U_i}=\sum_{l=1}^N\frac{r_{il}}{\rho-p_l}.
  \]
ares precisely the equations  \eqref{compatibility.rp.2.1} and
\eqref{compatibility.rp.3.1}, see \cite{otros}.

Now, the remaining  results follow from the equations
\begin{align*}
  r_{il}\frac{\partial p_l}{\partial U_j}-
  r_{jl}\frac{\partial p_l}{\partial
  U_i}&=\Big(  \rho_{il}\frac{ \partial \pi_l}{\partial U_j}-
   \rho_{jl}\frac{\partial  \pi_l}{\partial U_i}\Big)
   \pi_l-\rho_{il}R_{j,0}+
\rho_{jl}R_{i,0}\\
    \sum_{k\neq l}\frac{r_{ik}r_{jl}-r_{il}r_{jk}}{p_k-p_l}  &=\sum_{k\neq l}(\rho_{ik}\rho_{jl}-\rho_{il}\rho_{jk})
\pi_l\Big(\frac{ \pi_l}{\pi_k-\pi_l}+1 \Big)
\\
\frac{\partial r_{il}}{\partial U_j}-
 \frac{\partial  r_{jl}}{\partial
  U_i}&=\Big(\frac{\partial\rho_{il}}{\partial U_j}-
 \frac{\partial  \rho_{jl}}{\partial
  U_i}\Big)\pi_l  +\rho_{il}\frac{ \partial \pi_l}{\partial U_j}-
   \rho_{jl}\frac{\partial  \pi_l}{\partial U_i}
   \\
  2\sum_{k\neq
l}\frac{r_{ik}r_{jl}-r_{il}r_{jk}}{(p_l-p_k)^2} &=\sum_{k\neq
l}(\rho_{ik}\rho_{jl}-\rho_{il}\rho_{jk})\Big(
\frac{(\pi_k+\pi_l)\pi_l}{(\pi_k-\pi_l)^2}+\frac{
\pi_l}{\pi_k-\pi_l} \Big),
\end{align*}

  \end{enumerate}
\end{proof}

The relations\eqref{relations.KP},  \eqref{relations.dym},
\eqref{potential1.dym} and \eqref{potential2.dym} allow us to take
\[
u_0=\rho.
\]
and
\[
\begin{aligned}
  v&=\exp(\tilde\sigma),\\
  v_0&=\tilde\rho\exp(\tilde\sigma).
\end{aligned}
\]

Diagonal reductions appear when $\rho_{ij}=\rho_i\delta_{ij}$ and
$\tilde\rho_{ij}=\tilde\rho_i\delta_{ij}$ then
\[
R_i=\frac{p\rho_i}{p-\pi_i},\quad \tilde
R_i=\frac{p^2\tilde\rho_i}{p-\tilde\pi_i},
\]
and \eqref{compatibility.rp.0.2} and \eqref{compatibility.rp.0.2}
become
\begin{subequations}
 \begin{align}
 \label{compatibility.rp.0.2.1}
  \frac{\partial \pi_i}{\partial U_j}-
    &=\frac{\rho_{j}}{\pi_i-\pi_j}\pi_i,\\
  \label{compatibility.rp.0.3.1}
\frac{\partial \rho_{i}}{\partial U_j}
  &=-\frac{\rho_{i}\rho_{j}}{(\pi_i-\pi_j)^2}(\pi_j+\pi_i).
\end{align}
\end{subequations}
while \eqref{compatibility.dym.0.2} and
\eqref{compatibility.dym.0.3} read
\begin{subequations}
 \begin{align}
 \label{compatibility.dym.0.2.1}
  \frac{\partial \tilde\pi_i}{\partial U_j}-
    &=\frac{\tilde\rho_{j}}{\tilde\pi_i-\tilde\pi_j}\tilde\pi_i^2,\\
  \label{compatibility.dym.0.3.1}
\frac{\partial \tilde\rho_{i}}{\partial U_j}
  &=-2\frac{\tilde\rho_{i}\tilde\rho_{j}}{(\tilde\pi_i-\tilde\pi_j)^2}\tilde\pi_j\tilde\pi_i.
\end{align}
\end{subequations}

\subsection{Reductions for the $r$-dmKP hierarchy}

Here we shall consider the reduction \eqref{reduction.L} for
$R_i$ as in \eqref{R.form.0}. We also consider functions
$s_<\in\g_<$ satisfying
\begin{gather}\label{s<} \frac{\partial s_<}{\partial
p}R_i+\frac{\partial s_<}{\partial
  U_i}=p^{1-r}\sum_{j=1}^N
\frac{\rho_{ij}f_j}{p-p_j}
\end{gather}
where we suppose that the compatibility conditions for \eqref{s<}
\[
\rho_{il}\frac{\partial f_l}{\partial U_j}-\rho_{jl}\frac{\partial
f_l}{\partial U_i}=-\sum_{k\neq
l}(\rho_{ik}\rho_{jl}-\rho_{il}\rho_{jk})\Big(r+\frac{\pi_k}{\pi_l-\pi_k}\Big)\frac{f_l-f_k}{\pi_l-\pi_k}
\]
hold. If we deal with a diagonal reduction of the type
$\rho_{ij}=\rho_i\delta_{ij}$ the above compatibility conditions
become
\begin{gather}\label{comp.f.diag.kp}
\frac{\partial f_i}{\partial
U_j}=\frac{\rho_j}{\pi_i-\pi_j}\Big(r+\frac{\pi_j}{\pi_i-\pi_j}\Big)(f_i-f_j).
\end{gather}

Then, we have
\begin{pro}\label{Pro-KP}
  Let $s_<(p,\bu)\in\g_<$ be a function satisfying \eqref{s<} and define
  \begin{align*}
    s_\geq(p,\bu,\bt)&:=\sum_{n=1}^\infty
  t_n\Omega_n(p,\bu)\in\g_\geq,\\
  s(p,\bu,x,\bt)&:=s_\geq(p,\bu,\bt)+\Pi_r(p)x+s_<(p,\bu).
  \end{align*}
Suppose that $\bu=\bu(x,\bt)$ is determined by the following s
hodograph system
\begin{gather}\label{hodograph}
\sum_{n=1}^\infty t_n\frac{\partial \Omega_n}{\partial
p}(\pi_i(\bu),\bu)+x\pi_i(\bu)^{-r} +\pi_i(\bu)^{-r}f_i(\bu)=0,
\quad i=1,\dots,N.
\end{gather}
Then,
\[
S(L,x,\bt):=s(p(L,\bu),\bu,x,\bt)
\]
is an $S$-function.
\end{pro}
\begin{proof}
As we have
\begin{align*}
 \frac{\partial S}{\partial
t_n}&=\omega_n(L,\bu)+\sum_{i=1}^N\frac{\partial
s(p(L,\bu),\bu,x,\bt)}{\partial
U_i}\bigg|_{\bu=\bu(\bt)}\frac{\partial U_i}{\partial t_n},\\
 \frac{\partial S}{\partial x
}&=\Pi_r+\sum_{i=1}^N\frac{\partial
s(p(L,\bu),\bu,x,\bt)}{\partial
U_i}\bigg|_{\bu=\bu(\bt)}\frac{\partial U_i}{\partial x},
\end{align*}
the stated result follows from the identity
\[
\frac{\partial s(p(L,\bu),\bu,x,\bt)}{\partial
U_i}\bigg|_{\bu=\bu(\bt)}=0,
\]
that we shall show to hold. For this aim we first observe that
\begin{gather}\label{func.s}
\frac{\partial s(p(L,\bu),\bu,x,\bt)}{\partial U_i}=
\frac{\partial s}{\partial p}R_i+\frac{\partial s}{\partial U_i}.
\end{gather}
Then, multiplying  \eqref{s<} by $\prod_{l=1}^N(p-p_l)$, recalling
that $s_<$ is regular at $p=\pi_i$, and taking the limit $p\to
\pi_i$ we get
\[
\frac{\partial s_<}{\partial p}\bigg|_{p=\pi_i}=\pi_i^{-r}f_i,
\]
which together with \eqref{hodograph} implies
\begin{gather}\label{s_0}
  \frac{\partial s}{\partial p}\bigg|_{p=\pi_i}=0.
\end{gather}

Now, observing
\[
\frac{\partial s(p(L,\bu),\bu,x,\bt)}{\partial
U_i}=\sum_{n=1}^\infty t_n\frac{\partial\omega_n(L,\bu)}{\partial
U_i}+xp^{-r}R_i+\frac{\partial s_<}{\partial U_i}
\]
and recalling
\begin{align*}
\omega_n(L,\bu):=\Omega_n(p(L,\bu),\bu)=L^{n+1-r}-P_<(L^{n+1-r})&\Rightarrow\frac{\partial\omega_n(L,\bu)}{\partial
U_i}\in\g_< , \\
R_i=O(1),\quad p\to\infty&\Rightarrow p^{-r}R_i\in\g_<,
\end{align*}
we conclude
\[
\frac{\partial s(p(L,\bu),\bu,\bt)}{\partial U_i}\in\g_<,
\]
that, when applied to equation \eqref{func.s}, gives
\begin{gather}\label{s_1}
\frac{\partial s(p(L,\bu),\bu,x,\bt)}{\partial
U_i}=P_<\Big(\frac{\partial s}{\partial p}R_i\Big)+\frac{\partial
s_<}{\partial U_i}.\end{gather}

Let us introduce a function $E=E(p,\bu)\in\g_\geq$ such that
\begin{gather}\label{E}
  \frac{\partial E}{\partial p}(\pi_i,\bu)=\pi_i^{-r}f_i(\bu),
\end{gather}
for example we may take
\[
 \frac{\partial E}{\partial p}=p^{-r}\sum_{i=1}^Nf_i\prod_{j\neq
 i}\frac{p-\pi_j}{\pi_i-\pi_j}.
\]
Then, denoting $\hat s_\geq:=s_\geq+\Pi_r x$,
\[
P_<\Big(\frac{\partial s}{\partial
p}R_i\Big)=P_<\Big(\frac{\partial (\hat s_\geq+E)}{\partial
p}R_i\Big)+P_<\Big(\frac{\partial (s_<-E)}{\partial p}R_i\Big).
\]
In the one hand we notice that from \eqref{s_0} and
\eqref{hodograph} we have
\[ \frac{\partial
(\hat s_\geq+E)}{\partial
p}=p^{-r}\bigg(\prod_{i=1}^N(p-p_i)\bigg)(\alpha_0+\alpha_1p+\cdots)
\]
and hence
\[
\frac{\partial (\hat s_\geq+E)}{\partial
p}R_i=p^{1-r}(\alpha_0+\alpha_1p+\cdots)\sum_{j=1}^N\rho_{ij}\bigg(\prod_{\substack{l=1,\dots,N
\\
l\neq j}}(p-p_l)\bigg)\in\g_\geq,
\]
so that
\[
P_<\Big(\frac{\partial (\hat s_\geq+E)}{\partial p}R_i\Big)=0.
\]
In the other hand, we formula
\[
P_<\Big(\frac{\partial s_<}{\partial p}R_i\Big)=\frac{\partial
s_<}{\partial p}R_i
\]
which follows from $R_i=O(1)$ when $p\to \infty$ and
$\dfrac{\partial s_<}{\partial p}\in\g_<$, and we have the
relation
\[
P_<\Big(\frac{\partial E}{\partial p}R_i\Big)=p^{1-r}\sum_{j=1}^N
\frac{\rho_{ij}f_j}{p-p_j}.
\]

 Therefore,
\[
P_<\Big(\frac{\partial (s_<-E)}{\partial p}R_i\Big)=\frac{\partial
s_<}{\partial p}R_i-p^{1-r}\sum_{j=1}^N
\frac{\rho_{ij}f_j}{p-p_j}.
\]

Coming back to \eqref{s_1} we get
\[
\frac{\partial s(p(L,\bu),\bu,x,\bt)}{\partial U_i}=\frac{\partial
s_<}{\partial p}R_i+\frac{\partial s_<}{\partial
U_i}-p^{1-r}\sum_{j=1}^N \frac{\rho_{ij}f_j}{p-p_j}
\]
which vanishes in virtue of \eqref{s<}.
\end{proof}
\subsection{Hydrodynamic type systems and the $r$-dmKP
hierarchy}\label{ht}

Here we will briefly discuss how the reduction scheme derived from
\eqref{reduction.L} is associated with hydrodynamic type systems.
First, we remark that, assuming $L$ to be regular at the points
$p=\pi_i$, $i=1,\dots,N$, we have
\[
\frac{\partial L}{\partial p}\bigg|_{p=\pi_i}=0,
\]
so that \eqref{Lax.eqs} implies
\[
\sum_{j=1}^N\ell_{ij}\frac{\partial U_j}{\partial
t_n}=D_{in}\sum_{j=1}^N\ell_{ij}\frac{\partial U_j}{\partial x}
\]
where
 \[
L_i:=L\big|_{p=\pi_i},\quad D_{in}:=\pi_i^r\frac{\partial
\Omega_n}{\partial p}\bigg|_{p=\pi_i},\quad
\ell_{ij}:=\frac{\partial L_i}{\partial U_j}.
\]
Thus, if we define the following matrices
\[
D_n:=\text{diag}(D_{1n},\dots,D_{Nn}),\quad
\ell:=(\ell_{ij}),\quad A_n:=\ell^{-1}D_n\ell,
\]
we have the following hydrodynamic type system
\[
\frac{\partial \bu}{\partial t_n}=A_n(\bu)\frac{\partial
\bu}{\partial x}.
\]

Let us study in more detail the $t_1$-flow, as we know
\[
\Omega_1=p^{2-r}+(2-r)u_0p^{1-r}
\]
so that
\begin{gather}\label{1}
\frac{\partial L}{\partial
t_1}=(2-r)\Big(\big(p+(1-r)u_0\big)\frac{\partial L}{\partial
x}-p\frac{\partial u_0}{\partial x}\frac{\partial L}{\partial
p}\Big)
\end{gather}
that implies
\[
\frac{\partial u_n}{\partial t_1}=(2-r)\Big(\frac{\partial
u_{n+1}}{\partial x}+(1-r)u_0\frac{\partial u_n}{\partial
x}+nu_n\frac{\partial u_0}{\partial x}\Big),\quad n\geq 0,
\]
which we may think of as an $r$-modified Benney moment equations
---recall that in the dispersionless KP hierarchy the first
non-trivial flow comprimiese precisely the Benney moment
equations---.

If we define
\[
\Delta_n:=p^r\frac{\partial \Omega_n}{\partial p}\Rightarrow
\Delta_1=(2-r)(p+(1-r)u_0)
\]
we may write
\[
A_n=\Delta_n(\hat A_1),\quad \hat A_1:=\frac{A_1}{2-r}-(1-r)u_0
\]
which is equivalent to the Kodama--Gibbons formula for the
dispersionless KP equation \cite{kodama-gibbons}. The relevance of
the $r$-modified Benney equations also appears in relation with
the reduction \eqref{reduction.L}. If we introduce the reduction
in \eqref{1} we get
\[
\sum_{j=1}^N\bigg(\sum_{i=1}^N\frac{\partial L}{\partial U_i}(\hat
A_{1,ij}-p\delta_{ij})+p\frac{\partial u_0}{\partial
U_j}\frac{\partial L}{\partial p}\bigg)\frac{\partial
U_j}{\partial x}=0,
\]
and assuming the linear independence of $\dfrac{\partial
U_j}{\partial x}$, $j=1,\dots,N$, we have
\[
R_i=p\sum_{j=1}^N (\hat A_1-p)^{-1}_{ji}\frac{\partial
u_0}{\partial U_j}.
\]

\subsection{Examples of hodograph solutions of the $r$-dmKP hierarchy}
We analyze here some solutions derived, from Proposition
\ref{Pro-KP}, of the $r$-dmKP system \eqref{dmkp-system} ---which
implies the $r$-dmKP equation \eqref{r-dmKP}---. As we are
interested in \eqref{dmkp-system} we shall set $t_n=0$ for
$n=3,4,\dots$. For the $\Omega$'s we have
\begin{align*}
  \Omega_1&=p^{2-r}+(2-r)u_0p^{1-r},\\
  \Omega_2&=p^{3-r}+(3-r)u_0p^{2-r}+(3-r)\Big(u_1+\frac{2-r}{2}u_0^2\Big)
  p^{1-r}.
\end{align*}

\subsubsection{1-component reduction}

For $N=1$ there are no compatibility conditions to fulfill and
therefore we may take
\[
R_1=-\frac{p}{p-\pi_1(U)},\quad U=U_1.
\]
Equations \eqref{relations.KP} are
\begin{gather*}
  \frac{\partial u_0}{\partial U}=1,\quad
  \frac{\partial u_1}{\partial U}=\pi_1,\quad
    \frac{\partial u_2}{\partial U}=\pi_1^2-u_1,
\end{gather*}
so that we may take
\begin{gather}\label{uU}
  u_0=U,\quad u_1=\int^U\pi_1(s)\d s,\quad u_2=\int^U
  \pi_1(s)^2\d s-\int^U(\int^s \pi_1(s')\d s')\d s.
\end{gather}

In this situation $\hat s_\geq$ is
\[
\hat
s_\geq=\Big(p^{3-r}+(3-r)u_0p^{2-r}+(3-r)\Big(u_1+\frac{2-r}{2}u_0^2\Big)
p^{1-r}\Big)t_2+(p^{2-r}+(2-r)u_0p^{1-r})t_1+\Pi_rx.
\]
Therefore, the hodograph condition \eqref{hodograph} is
\[
(3-r)\Big(\pi_1^2+(2-r)u_0\pi_1+(1-r)\Big(u_1+\frac{2-r}{2}u_0^2\Big)\Big)t_2+
(2-r)(\pi_1+(1-r)u_0)t_1+x+f(U)=0
\]
where $f$ is an arbitrary function. If we now introduce formulae
\eqref{uU} we get the following
\begin{pro}
 Given two arbitrary functions $\pi_1(U),f(U)$, and a function  $U(x,t_1,t_2)$
 determined by the hodograph equation
 \begin{multline}
  \label{hodograph-mkp-1}
(3-r)\Big(\pi_1(U)^2+(2-r)U\pi_1(U)+(1-r)\Big(\int^U\pi_1(s)\d
s+\frac{2-r}{2}U^2\Big)\Big)t_2\\
+ (2-r)(\pi_1(U)+(1-r)U)t_1+x+f(U)=0.
\end{multline}
then
\begin{gather}\label{uU2}
  u_0=U,\quad u_1=\int^U\pi_1(s)\d s,\quad u_2=\int^U
  \pi_1(s)^2\d s-\int^U\int^s \pi_1(s')\d s'\d s.
\end{gather}
solves the $r$-dmKP system \eqref{dmkp-system}.
\end{pro}
 While this is a
consequence of Proposition \ref{Pro-KP} the following direct proof
is available.
\begin{proof} If
we request to  $u_0,u_1$ and $u_2$ as in \eqref{uU} to solve the
system of PDE's \eqref{dmkp-system} then $U(x,t_1,t_2)$  satisfy
\begin{gather}\label{hodo-diff}
\begin{aligned}
  \frac{U_{t_1}}{U_x}&=(2-r)(\pi_1(U)+(1-r)U),\\
\frac{U_{t_2}}{U_x}&=(3-r)\Big(\pi_1(U)^2+(2-r)U\pi_1(U)+(1-r)\Big(\int^U\pi_1(s)\d
s+\frac{2-r}{2}U^2\Big)\Big).
\end{aligned}
\end{gather}
But, taking $x$, $t_1$ and $t_2$ derivatives in
\eqref{hodograph-mkp-1} we get
\begin{align*}
  (f'(U)+B'(U)t_2+A'(U)t_1)U_x+1&=0,\\
  (f'(U)+B'(U)t_2+A'(U)t_1)U_{t_1}+A(U)&=0,\\
  (f'(U)+B'(U)t_2+A'(U)t_1)U_{t_2}+B(U)&=0,
\end{align*}
where
\begin{align*}
  A&:=(2-r)(\pi_1(U)+(1-r)U),\\
  B&:=(3-r)\bigg(\pi_1(U)^2+(2-r)U\pi_1(U)+(1-r)\Big(\int^U\pi_1(s)\d
s+\frac{2-r}{2}U^2\Big)\bigg),
\end{align*}
which imply \eqref{hodo-diff}. Thus, solutions $U(x,t_1,t_2)$ of
\eqref{hodograph-mkp-1} provide us with solutions to
\eqref{dmkp-system}.
\end{proof}

A simple solution appears with the choice $\pi_1=kU$, $f:=0$.
Observe that  we are dealing, as follows from \eqref{reduction.L},
with the following reduction
\[
L=\begin{cases}   p\,\Exp{Up^{-1}}& k=1,\\
p\,(1+(1-k)Up^{-1})^{\frac{1}{1-k}},& k\neq 1.
\end{cases}
\]
If this is the case we get
\[
A=\alpha U\quad B:=\beta U^2
\]
with
\[
\alpha:=(2-r)(k+1-r),\quad\beta:=(3-r)\Big(k^2+(5-3r)k+\frac{(1-r)(2-r)}{2}\Big)
\]
and the corresponding solution, for $\beta\neq 0$, is
\begin{gather}
  \label{k}u_0=U=-\frac{\alpha t_1}{2\beta t_2}\pm\sqrt{\frac{\alpha^2
t_1^2}{4\beta^2 t_2^2}-\frac{x}{\beta t_2}}.
\end{gather}

There are two particular values for $k$
\[
k=-(2-r),-\frac{1-r}{2}
\]
such that
\[
\beta=0 \text{ and $\alpha=-(2-r)$ and $\dfrac{(2-r)(1-r)}{2}$,
respectively}.
\]
In this case  we get two simple $t_2$-invariant solutions
\[
u_0=\begin{cases} -\dfrac{x}{(2-r)t_1}&\text{ for $k=r-2$,}\\[8pt]
\dfrac{2x}{(1-r)(2-r)t_1}&\text{ for $k=(r-1)/2$.}
\end{cases}
\]

For example, if in instead of $f=0$ we set $f=U^3$ we get the
solution
\[
  u_0=U=g-\frac{3\alpha t_1-\beta^2t_2^2}{9g}-\frac{\beta t_2}{3}
  \]
where $g$ is defined by
  \begin{multline*}
g:=\frac{1}{6}\sqrt[3]{-108x+36\alpha\beta t_1t_2-8\beta^2t_2^2
+12\sqrt{81 x^2+12\alpha^3t_1^3-54\alpha\beta
xt_1t_2+12\beta^3xt_2^3-3\alpha^2\beta^2t_1^2t_2^2}}
\end{multline*}
For $\beta=0$ we get the following $t_2$-independent solution
\[
  u_0=U=g-\frac{3\alpha t_1}{9g},\text{ where }g:=\frac{1}{6}\sqrt[3]{-108x +12\sqrt{81 x^2+12\alpha^3t_1^3}}
  \text{ for
$\alpha=-(2-r),\dfrac{(2-r)(1-r)}{2}$.}
  \]

\subsubsection{2-component reduction}

We shall work with the diagonal reduction given by
\eqref{compatibility.rp.0.2.1} and \eqref{compatibility.rp.0.3.1}.
We may take the following solution  for $N=2$:
\[
\pi_1=-\pi_2=\frac{U_1-U_2}{4},\quad\rho_1=\rho_2=-\frac{1}{2}.
\]
The linear system \eqref{comp.f.diag.kp} for $f_1,f_2$ becomes
\begin{gather}\label{f.two.c}
  \frac{\partial f_1}{\partial
  U_2}=\frac{\partial f_2}{\partial
  U_1}=-\frac{2r-1}{2(U_1-U_2)}(f_1-f_2),
\end{gather} which is equivalent to
\[
f_1=\frac{\partial \Phi}{\partial U_1},\quad
f_2=\frac{\partial\Phi}{\partial U_2},
\]
with
\[
\frac{\partial^2 \Phi}{\partial U_1\partial
U_2}+\frac{2r-1}{2(U_1-U_2)}\Big(\frac{\partial \Phi}{\partial
U_1}-\frac{\partial \Phi}{\partial U_2}\Big)=0.
\]
The method of separation of variables leads to the following
solutions, expressed in terms of the Bessel and Neumann functions
$J_{-r},N_{-r}$ (the Neumann function is also known as the Weber
function $Y_{-r}$), for a similar result for the dKP hierarchy see
\cite{otros}:
\[
\Phi=(U_1-U_2)^r\big(A J_{-r}(k (U_1-U_2))+B N_{-r}(k
(U_1-U_2))\big)\big(C\cos(k (U_1+U_2))+D\sin(k (U_1+U_2))\big)
\]
and also
\[
\Phi=\begin{cases}   (A+B(U_1-U_2)^{2r})(C+D(U_1+U_2)), &r\neq
0,\\
(A+B\log(U_1-U_2))(C+D(U_1+U_2))& r=0.
\end{cases}
\]

From
\begin{align*}
  \frac{\partial u_0}{\partial U_i}&=\frac{1}{2},\\
\frac{\partial u_1}{\partial U_i}&=\frac{1}{2}\pi_i,\\
\frac{\partial u_2}{\partial U_i}&=\frac{1}{2}(\pi_i^2-u_1),
\end{align*}
for $i=1,2$, we get
\begin{gather*}
  u_0=\frac{U_1+U_2}{2},\quad u_1=\frac{(U_1-U_2)^2}{16},\quad
  u_2=0.
\end{gather*}

From the formulae
\begin{align*}
  \pi_i^2+(2-r)u_0\pi_i+(1-r)\Big(u_1+\frac{2-r}{2}u_0^2\Big)&=\begin{cases}
    \dfrac{2-r}{2}\Big((1-r)U_+^2+\dfrac{1}{2}U_-^2+U_+U_-\Big),& i=1,\\[6pt]
 \dfrac{2-r}{2}\Big((1-r)U_+^2+\dfrac{1}{2}U_-^2-U_+U_-\Big),&i=2,
  \end{cases}\\
\pi_i+(1-r)u_0&=\begin{cases}
(1-r)U_++\dfrac{1}{2}U_-,&i=1,\\[6pt]
(1-r)U_+-\dfrac{1}{2}U_-,&i=2,
\end{cases}
\end{align*}
where
\[
U_\pm:=\frac{U_1\pm U_2}{2}
\]
we deduce the following hodograph system
\begin{gather}\label{sys_hodo_2}
  \begin{aligned}
\dfrac{(3-r)(2-r)}{2}\Big((1-r)U_+^2+\dfrac{1}{2}U_-^2+U_+U_-\Big)t_2
+(2-r)\Big((1-r)U_++\dfrac{1}{2}U_-\Big)t_1+x&
=f_1,\\
\dfrac{(3-r)(2-r)}{2}\Big((1-r)U_+^2+\dfrac{1}{2}U_-^2-U_+U_-\Big)t_2
+(2-r)\Big((1-r)U_+-\dfrac{1}{2}U_-\Big)t_1+x& =f_2.
\end{aligned}
\end{gather}

Adding and subtracting the equations of \eqref{sys_hodo_2} we
obtain the equivalent system
\begin{gather}\label{sys_hodo_2.2}
  \begin{gathered}
\frac{(3-r)(2-r)}{2}\Big((1-r)U_+^2+\dfrac{U_-^2}{2}\Big)
t_2+(2-r)(1-r)U_+t_1+x
=\frac{f_1+f_2}{2},\\
(3-r)(2-r)U_+U_-t_2+(2-r)U_-t_1=f_1-f_2.
\end{gathered}
\end{gather}
The simplest solution of \eqref{f.two.c} is $f_1=f_2=0$; i.e.,
$\Phi=0$, in this case there are two possible choices
\[
U_-=0 \text{ or } U_+=-\frac{1}{3-r}\frac{t_1}{t_2}.
\]
The first choice imply $u_1=U_-^2/4=0$ and $u_0=U_+$ to solve the
algebraic equation
\[
\frac{(3-r)(2-r)(1-r)}{2}t_2U_+^2 +(2-r)(1-r)t_1U_++x=0
\]
whose solutions are
\[
u_0=U_+=-\frac{t_1}{(3-r)t_2}\pm\sqrt{\frac{t_1^2}{(3-r)^2t_2^2}-\frac{2x}{(3-r)(2-r)(1-r)t_2}}.
\]
In fact, for $u_1=u_2=0$ the system \eqref{dmkp-system} becomes
\begin{gather*}
\begin{aligned}
u_{0,t_1}&=(2-r)(1-r)u_0u_{0,x},\\
u_{0,t_2}&=\frac{1}{2}(3-r)(2-r)(1-r)u_0^2u_{0,x};
\end{aligned}
\end{gather*}
i.e., the dispersionless KdV ($t_1$-flow) and dispersionless
modified KdV ($t_2$-flow) equations, and $u_0=U_+$ is a solution
of both simultaneously. This solution is a particular case of the
already studied 1-component solution \eqref{k}; i.e $\pi_1=kU$,
for $k=0$ but this value of $k$ is not allowed as we assume that
$\pi_1\neq 0$, therefore the two-component case, with
$\pi_1=-\pi_2=(U_1-U_2)/4$, provides us with a rigorous
alternative track to this solution.

The second choice
\[
u_0=U_+=-\frac{1}{3-r}\frac{t_1}{t_2}
\]
imply
\[
u_1=\frac{U_-^2}{4}=\frac{(1-r)t_1^2}{2(3-r)^2}\frac{t_1^2}{t_2^2}-\frac{1}{(3-r)(2-r)}\frac{x}{t_2}.
\]
It is easy to check that this pair $u_0,u_1$ fulfills the $u_2=0$
reduction of  \eqref{dmkp-system}, namely
\begin{gather*}
\begin{aligned}
u_{0,t_1}&=(2-r)u_{1,x}+(2-r)(1-r)u_0u_{0,x},\\
u_{1,t_1}&=(2-r)(1-r)u_0u_{1,x}+(2-r)u_1u_{0,x},\\
u_{0,t_2}&=(3-r)(2-r)(u_0u_{1,x}+u_{0,x}u_1)+\frac{1}{2}(3-r)(2-r)(1-r)u_0^2u_{0,x}.
\end{aligned}
\end{gather*}

 A more general choice is to take
$\Phi=\Phi_+(U_+)\Phi_-(U_-)$, then
\begin{align*}
f_1&=\dfrac{1}{2}(\Phi_+'(U_+)\Phi_-(U_-)  +\Phi_+(U_+)\Phi_-'(U_-)),\\
f_2&=\dfrac{1}{2}(\Phi_+'(U_+)\Phi_-(U_-)-\Phi_+(U_+)\Phi_-'(U_-)),
\end{align*}
and, if we assume $\Phi'_-\neq 0$, the hodograph system
\eqref{sys_hodo_2.2} reads
\begin{gather}\label{sys_hodo_2.3}
\begin{gathered}
\frac{(3-r)(2-r)}{2}\Big((1-r)U_+^2+\dfrac{U_-^2}{2}\Big)
t_2+(2-r)(1-r)U_+t_1+x
=\frac{\Phi_+'(U_+)\Phi_-(U_-)}{4},\\
U_-((3-r)(2-r)U_+t_2+(2-r)t_1)=\frac{\Phi_+(U_+)\Phi_-'(U_-)}{2}.
\end{gathered}
\end{gather}

Picking, for example, $\Phi_-=U_-^{2r}$, $\Phi_+=A+BU_+$  we have
for $u_1=U_-^2/4$
\[
u_1=\frac{1}{4}U_-^{2}
=\frac{1}{4}\Big(\frac{r(A+BU_+)}{(2-r)t_1+(3-r)(2-r)t_2U_+}\Big)^{\frac{1}{1-r}}
\]
and $u_0=U_+$ is implicitly determined by
\[
\frac{(3-r)(2-r)(1-r)}{2}t_2U_+^2+(2-r)(1-r)t_1U_++x=\frac{B}{4}(4u_1)^r-(3-r)(2-r)t_2u_1
.
\]

\subsection{Reductions for the $r$-dDym hierarchy}

Here we tackle the reduction \eqref{reduction.tilde.L}  with
$\tilde R_i$ as in \eqref{tilde.R.form.0} for the $r$-dDym
hierarchy. Let us introduce $s_\leq\in\g_\leq$ satisfying
\begin{gather}\label{s<.dym} \frac{\partial \tilde s_\leq}{\partial
p}\tilde R_i+\frac{\partial \tilde s_\leq}{\partial
  U_i}=p^{2-r}\sum_{j=1}^N
\frac{\tilde\rho_{ij}\tilde f_j}{p-\tilde\pi_j}.
\end{gather}
Here we assume the  compatibility conditions for \eqref{s<.dym}
\[
\tilde\rho_{il}\frac{\partial \tilde f_l}{\partial
U_j}-\tilde\rho_{jl}\frac{\partial\tilde f_l}{\partial
U_i}=-\sum_{k\neq
l}(\tilde\rho_{ik}\tilde\rho_{jl}-\tilde\rho_{il}\tilde\rho_{jk})\Big(-(1-r)
+\frac{\tilde
\pi_k}{\tilde\pi_l-\tilde\pi_k}\Big)\tilde\pi_l\frac{\tilde
f_l-\tilde f_k}{\tilde \pi_l-\tilde \pi_k},
\]
that for the diagonal reduction $\tilde\rho_{ij}=\tilde
\rho_i\delta_{ij}$ are
\begin{gather}
  \label{tilde-diag}
\frac{\partial \tilde f_i}{\partial U_j}=\Big(-(1-r) +\frac{\tilde
\pi_j}{\tilde\pi_i-\tilde\pi_j}\Big)\frac{\tilde\rho_{j}\tilde\pi_i}{\tilde
\pi_i-\tilde \pi_j}(\tilde f_i-\tilde f_j).\end{gather}

 We have
\begin{pro}\label{Pro-Dym}\label{tildes}
  Let $\tilde s_\leq(p,\bu)\in\g_\leq$ satisfying \eqref{s<.dym}
  and define
  \begin{gather*}
    \tilde s_>(p,\bu,\bt):=\sum_{n=1}^\infty
  t_n\tilde\Omega_n(p,\bu),\\
\tilde   s(p,\bu,x,\bt):=\tilde s_>(p,\bu,\bt)+\Pi_r(p)\,x+\tilde
s_\leq(p,\bu).
\end{gather*}
Suppose that $\bu=\bu(x,\bt)$ is determined by the following
hodograph system
\begin{gather}\label{tilde.hodograph}
\sum_{n=1}^\infty t_n\frac{\partial \tilde\Omega_n}{\partial
p}(\tilde\pi_i(\bu),\bu) +x\tilde\pi_i(\bu)^{-r}
+\tilde\pi_i(\bu)^{-r}\tilde f_i(\bu)=0, \quad i=1,\dots,N.
\end{gather}
Then,
\[
\tilde S(\tilde L,x,\bt):=\tilde s(p(\tilde L,\bu),\bu,x,\bt)
\]
is an $S$-function for the $r$-dDym hierarchy.
\end{pro}
\begin{proof}
The lines of this proof are almost identical to those followed in
Proposition \ref{Pro-KP} for the $r$-dmKP hierarchy. First we
observe that
\begin{align*}
 \frac{\partial \tilde S}{\partial t_n}=\tilde \omega_n(\tilde
L,\bu)+\sum_{i=1}^N\frac{\partial \tilde s(p(\tilde
L,\bu),\bu,x,\bt)}{\partial
U_i}\bigg|_{\bu=\bu(\bt)}\frac{\partial
U_i}{\partial t_n},\\
\frac{\partial \tilde S}{\partial
x}=\Pi_r+\sum_{i=1}^N\frac{\partial \tilde s(p(\tilde
L,\bu),\bu,x,\bt)}{\partial
U_i}\bigg|_{\bu=\bu(\bt)}\frac{\partial U_i}{\partial x},
\end{align*}
and our result will follow  if the formula
\[ \frac{\partial \tilde
s(p(\tilde L,\bu),\bu,x,\bt)}{\partial U_i}\bigg|_{\bu=\bu(\bt)}=0
\]
is satisfied.
 We will proceed observing that
\begin{gather}\label{func.tilde.s}
\frac{\partial \tilde s(p(\tilde L,\bu),\bu,x,\bt)}{\partial U_i}=
\frac{\partial\tilde  s}{\partial p}\tilde R_i+\frac{\partial
\tilde s}{\partial U_i}.
\end{gather}
Let us multiply \eqref{s<.dym} by $\prod_{l=1}^N(p-\tilde \pi_l)$,
also recall that $\tilde s_\leq$ is regular at $p=\tilde \pi_i$,
and then take the limit $p\to \tilde \pi_i$ in order to get
\[
\frac{\partial \tilde s_\leq}{\partial p}\bigg|_{p=\tilde
\pi_i}=\tilde \pi_i^{-r}\tilde f_i,
\]
and  from \eqref{tilde.hodograph} we obtain
\begin{gather}\label{s_0.tilde}
  \frac{\partial \tilde s}{\partial p}\bigg|_{p=\tilde\pi_i}=0.
\end{gather}

We notice that
\[
\frac{\partial \tilde s(p(\tilde L,\bu),\bu,x,\bt)}{\partial
U_i}=\sum_{n=1}^\infty t_n\frac{\partial\tilde \omega_n(\tilde
L,\bu)}{\partial U_i}+ xp(\tilde L,\bu)^{-r}\frac{\partial\tilde
p(\tilde L,\bu)}{\partial U_i}+\frac{\partial \tilde
s_\leq}{\partial U_i},
\]
and also that
\begin{align*}
\tilde \omega_n(\tilde L,\bu):=\tilde \Omega_n(p(\tilde
L,\bu),\bu)=\tilde L^{n+1-r}-P_\leq(\tilde
L^{n+1-r})&\Rightarrow\frac{\partial\tilde \omega_n(\tilde
L,\bu)}{\partial
U_i}\in\g_\leq ,\\
\text{ and if $r\neq 1$ }p(\tilde
L,\bu)^{1-r}\in\g_\leq\Rightarrow p(\tilde
L,\bu)^{-r}\frac{\partial p(\tilde L,\bu)}{\partial
U_i}\in\g_\leq,\\
\text{finally, for $r=1$, } \frac{\partial p}{\partial
U_i}=O(p)\text{ when $p\to\infty$ }\Rightarrow
p^{-1}\frac{\partial p}{\partial U_i}\in\g_\leq.
\end{align*}
Therefore, we can write
\[
\frac{\partial \tilde s(p(\tilde L,\bu),\bu,\bt)}{\partial
U_i}\in\g_\leq
\]
and from \eqref{func.tilde.s} we deduce that
\begin{gather}\label{s_1.tilde}
\frac{\partial \tilde s(p(\tilde L,\bu),\bu,x,\bt)}{\partial
U_i}=P_\leq\Big(\frac{\partial \tilde s}{\partial p}\tilde
R_i\Big)+\frac{\partial \tilde s_\leq}{\partial U_i}.
\end{gather}

Following the $r$-dmKP case we now introduce a function $\tilde
E=\tilde E(p,\bu)\in\g_\geq$ such that
\begin{gather}\label{E}
  \frac{\partial \tilde  E}{\partial p}(\tilde \pi_i,\bu)=\tilde \pi_i^{-r}f_i(\bu),
\end{gather}
for example we may take
\[
 \frac{\partial\tilde  E}{\partial p}=p^{-r}\sum_{i=1}^N\tilde f_i\prod_{j\neq
 i}\frac{p-\tilde \pi_j}{\tilde \pi_i-\tilde \pi_j}.
\]
Let us study the following equation
\[
P_\leq\Big(\frac{\partial\tilde  s}{\partial p}\tilde
R_i\Big)=P_\leq\Big(\frac{\partial (\tilde s_>+x\Pi_r+\tilde
E)}{\partial p}\tilde R_i\Big)+P_\leq\Big(\frac{\partial (\tilde
s_\leq -\tilde E)}{\partial p}\tilde R_i\Big).
\]
For the first term in the r.h.s we see that, as follows  that from
\eqref{s_0.tilde},
\[ \frac{\partial
(\tilde s_>+x\Pi_r+\tilde E)}{\partial
p}=p^{-r}\bigg(\prod_{i=1}^N(p-\tilde
\pi_i)\bigg)(\alpha_0+\alpha_1p+\cdots),
\]
and hence
\[
\frac{\partial (\tilde s_>+x\Pi_r+\tilde E)}{\partial p}\tilde
R_i=p^{2-r}(\alpha_0+\alpha_1p+\cdots)\sum_{j=1}^N\tilde
\rho_{ij}\bigg(\prod_{\substack{l=1,\dots,N
\\
l\neq j}}(p-\tilde \pi_l)\bigg)\in\g_>,
\]
so that
\[
P_\leq\Big(\frac{\partial (\tilde s_>+x\Pi_r+E)}{\partial p}\tilde
R_i\Big)=0.
\]
We also deduce that
\[
P_\leq\Big(\frac{\partial \tilde s_\leq}{\partial p}\tilde
R_i\Big)=\frac{\partial \tilde s_\leq}{\partial p}\tilde R_i,
\]
which is a direct consequence of $\tilde R_i=O(p)$ when $p\to
\infty$ and $\dfrac{
\partial\tilde s_\leq}{\partial p}\in\g_\leq$.  Another relevant
formula is
\[
P_\leq\Big(\frac{\partial \tilde E}{\partial p}\tilde
R_i\Big)=p^{2-r}\sum_{j=1}^N \frac{\tilde \rho_{ij}\tilde
f_j}{p-\tilde \pi_j}.
\]
With all these formula at hand we deduce
\[
P_\leq\Big(\frac{\partial (\tilde s_\leq-\tilde E)}{\partial
p}\tilde R_i\Big)=\frac{\partial \tilde s_\leq}{\partial p}\tilde
R_i-p^{2-r}\sum_{j=1}^N \frac{\tilde \rho_{ij}\tilde f_j}{p-\tilde
\pi_j}.
\]

Coming back to \eqref{s_1.tilde} we get
\[
\frac{\partial\tilde  s(p(\tilde L,\bu),\bu,x,\bt)}{\partial
U_i}=\frac{\partial \tilde s_\leq}{\partial p}\tilde
R_i+\frac{\partial\tilde  s_\leq}{\partial
U_i}-p^{1-r}\sum_{j=1}^N \frac{\tilde \rho_{ij}\tilde
f_j}{p-\tilde \pi_j}
\]
which vanishes in virtue of \eqref{s<.dym}.
\end{proof}

\subsection{Hydrodynamic type systems and the $r$-dDym hierarchy}

Here we proceed as in \S \ref{ht}. 
With the use of \eqref{LLax.eqs}, together with $\dfrac{\partial
\tilde L}{\partial p}\bigg|_{p=\tilde \pi_i}=0$, and of  the
notation
 \begin{gather*}
  \tilde L_i:=\tilde L\big|_{p=\tilde\pi_i},\quad \tilde
D_{in}:=\tilde \pi_i^r\frac{\partial \tilde \Omega_n}{\partial
p}\bigg|_{p=\tilde\pi_i},\quad \tilde\ell_{ij}:=\frac{\partial
\tilde L_i}{\partial U_j},\\
\tilde D_n:=\text{diag}(\tilde D_{1n},\dots,\tilde D_{Nn}),\quad
\tilde \ell:=(\tilde \ell_{ij}),\quad \tilde A_n:=\tilde
\ell^{-1}\tilde D_n\tilde \ell.
 \end{gather*}
we derive the following hydrodynamic type system
\[
\frac{\partial \bu}{\partial t_n}=\tilde A_n(\bu)\frac{\partial
\bu}{\partial x}.
\]

To analyze the $t_1$-flow, we recall that
\[
\tilde\Omega_1=v^{2-r}p^{2-r}
\]
and therefore
\begin{gather}\label{tilde1}
\frac{\partial \tilde L}{\partial
t_1}=(2-r)v^{1-r}p\Big(v\frac{\partial\tilde L}{\partial
x}-p\frac{\partial v}{\partial x}\frac{\partial \tilde L}{\partial
p}\Big)
\end{gather}
that implies the following Benney moment type equations
\[
\frac{\partial v_n}{\partial t_1}=(2-r)v^{1-r}\Big(v\frac{\partial
v_{n+1}}{\partial x}+(n+1)\frac{\partial v}{\partial
x}v_{n+1}\Big), \quad n\geq -1,\quad v_{-1}:=v.
\]

We now define
\[
\tilde\Delta_n:=p^r\frac{\partial \tilde\Omega_n}{\partial
p}\Rightarrow \tilde\Delta_1=(2-r)v^{2-r}p
\]
so that the corresponding Kodama--Gibbons formula for the $r$-dDym
hierarchy is
\[
\tilde A_n=\tilde\Delta_n(\hat{\tilde A}_1),\quad  \hat{\tilde
A}_1:=\frac{\tilde A_1}{(2-r)v^{2-r}}.
\]
 If we introduce the reduction
in \eqref{tilde1} we get
\[
\sum_{j=1}^N\bigg(\sum_{i=1}^N\frac{\partial \tilde L}{\partial
U_i}(\hat{\tilde A}_{1,ij}-p\delta_{ij})+p^2v\frac{\partial
v}{\partial U_j}\frac{\partial\tilde L}{\partial
p}\bigg)\frac{\partial U_j}{\partial x}=0,
\]
and assuming the linear independence of $\dfrac{\partial
U_j}{\partial x}$, $j=1,\dots,N$, we have
\[
\tilde R_i=p^2\sum_{j=1}^N (\hat{\tilde
A}_1-p)^{-1}_{ji}v\frac{\partial v}{\partial U_j}.
\]

\subsection{Examples of hodograph solutions of the $r$-dDym hierarchy}
We now ready to present here some solutions of the $r$-dDym system
\eqref{dDym-system}, which implies the $r$-dDym equation
\eqref{r-dDym}. We set $t_n=0$ for $n=3,4,\dots$ and recall that
\begin{align*}
  \tilde\Omega_1&=v^{2-r}p^{2-r},\\
  \tilde\Omega_2&=v^{3-r}p^{3-r}+(3-r)v_0v^{2-r}p^{2-r}.
\end{align*}

\subsubsection{1-component reduction}

We take
\[
\tilde R_1=-\frac{p^2}{p-\tilde\pi_1(U)},\quad U=U_1.
\]
From equations \eqref{relations.dym} we deduce the following
equations
\begin{gather}\label{uU.dym.diff}
v_U=v,\quad v_{0,U}=\tilde\pi_1 v,\quad
v_{1,U}=\tilde\pi_1^2v-v_1,
\end{gather}
which are solve by
\begin{gather}\label{uU.dym}
v=\Exp{U},\quad v_0=\int^U\tilde\pi_1(s)\Exp{s}\d s,\quad
v_1=\Exp{-U}\int^U
  \tilde\pi_1(s)^2\Exp{2s}\d s.
\end{gather}

In this situation $\tilde s_>$ is
\[
\tilde
s_>=\big(v^{3-r}p^{3-r}+(3-r)v_0v^{2-r}p^{2-r}\big)t_2+(v^{2-r}p^{2-r})t_1.
\]
Therefore, the hodograph condition \eqref{tilde.hodograph} is
\[
(3-r)v^{2-r}(v\tilde \pi_1^2+(2-r)v_0\tilde \pi_1)t_2+ (2-r)\tilde
\pi_1v^{2-r}t_1+x+f(U)=0
\]
where $f$ is an arbitrary function, which taking into account
\eqref{uU.dym} leads to the following
\begin{pro}
Given arbitrary two functions $\tilde\pi_1(U),f(U)$  and a
function $U(x,t_1,t_2)$ determined by
\begin{multline} \label{hodograph-dym-1}
(3-r)\Exp{(2-r)U}\Big(\Exp{U}\tilde\pi_1(U)^2+(2-r)\tilde
\pi_1(U)\int^U\tilde\pi_1(s)\Exp{s}\d s\Big)t_2+ (2-r)\tilde
\pi_1(U)\Exp{(2-r)U}t_1+x+f(U)=0
\end{multline}
then
\begin{gather*}
v=\Exp{U},\quad v_0=\int^U\tilde\pi_1(s)\Exp{s}\d s,\quad
v_1=\Exp{-U}\int^U
  \tilde\pi_1(s)^2\Exp{2s}\d s.
\end{gather*}
solves the $r$-dDym system \eqref{dDym-system}.
\end{pro}

This result is a corollary of Proposition \ref{tildes}, however we
give a direct proof.

\begin{proof}
The imposition  to $v,v_0$ and $v_1$,  as in \eqref{uU.dym.diff},
to solve \eqref{dDym-system}  is equivalent  to request
$U(x,t_1,t_2)$ to fulfill
\begin{gather}\label{hodo-diff.dym}
\begin{aligned}
  \frac{U_{t_1}}{U_x}&=(2-r)v^{2-r}\tilde\pi_1,\\
\frac{U_{t_2}}{U_x}&=(3-r)v^{2-r}(\tilde\pi_1(U)^2v+(2-r)\tilde
\pi_1(U)v_0).
\end{aligned}
\end{gather}
But if we take $x$, $t_1$ and $t_2$ derivatives in
\eqref{hodograph-dym-1} we get
\begin{align*}
  (f'(U)+B'(U)t_2+A'(U)t_1)U_x+1&=0,\\
  (f'(U)+B'(U)t_2+A'(U)t_1)U_{t_1}+A(U)&=0,\\
  (f'(U)+B'(U)t_2+A'(U)t_1)U_{t_2}+B(U)&=0,
\end{align*}
where
\begin{align*}
  A&:=(2-r)v^{2-r}\tilde\pi_1,\\
  B&:=(3-r)v^{2-r}(\tilde\pi_1(U)^2v+(2-r)\tilde\pi_1(U)v_0),
\end{align*}
which imply \eqref{hodo-diff.dym} and the desired result follows.
\end{proof}

Let us suppose that $\tilde\pi_1=\Exp{kU}$, then
\eqref{hodograph-dym-1} is
\[
\frac{(3-r)(3-r+k)}{1+k}t_2\Exp{(3-r+2k)U}+
(2-r)t_1\Exp{(2-r+k)U}+x=f,
\]

If we now impose
\[
3-r+2k=n(2-r+k)\Rightarrow k=\frac{3-r-n(2-r)}{n-2}
\]
we have
\[
\frac{(3-r)(3-r+k)}{1+k}t_2\alpha^n+ (2-r)t_1\alpha+x=0,
\]
where
\[
\alpha:=\Exp{(2-r+k)U}.
\]

We now explore in more detail the following three  examples
\begin{center}
\setlength{\extrarowheight}{10pt}
\begin{tabular}[b]{|c|c|c|c|}
\hline
 $n$ & $k$                & Hodograph equation & $v=v(\alpha)$ \\
\hline
   3 & $-3+2r$            & $-\dfrac{(3-r)r}{2(1-r)}t_2\alpha^3+C\alpha^2+(2-r)t_1\alpha+x=0$  & $\alpha^{-\frac{1}{1-r}}$   \\
  $-1$ & $\dfrac{-5+2r}{3}$ & $(2-r)t_1\alpha^2+x\alpha-\dfrac{(3-r)(4-r)}{2(1-r)}t_2=0$ & $\alpha^{\frac{3}{1-r}}$  \\
  $-2$ & $\dfrac{-7+3r}{4}$ & $(2-r)t_1\alpha^3+x\alpha^2+C\alpha-\dfrac{(3-r)(5-r)}{3(1-r)}t_2=0$ &  $\alpha^{\frac{4}{1-r}}$\\
  \hline
\end{tabular}
\end{center}
where $C$ is an arbitrary constant which appears by choosing $f$
in an appropriate manner. The hodograph equations in this cases
are explicitly solved, as we are dealing with cubic and a
quadratic equations. Now, we present the corresponding solution
the $r$-dDym equation \eqref{r-dDym} in the $n=-1$ case:
\[
v=\left( -\dfrac{1}{2(2-r)}\dfrac{x}{t_1}\pm\sqrt{
\dfrac{1}{4(2-r)^2}\dfrac{x^2}{t_1^2}+\dfrac{(3-r)(4-r)}{2(1-r)(2-r)}\dfrac{t_2}{t_1}}\right)^{
\frac{3}{1-r}},
\]
for $n=3$, $C=0$, the corresponding solution is
\[
v=\left(-\frac{(1-r)g(x,t_1,t_2)}{(3-r)rt_2}-\frac{2(2-r)t_1}{g(x,t_1,t_2)}\right)^{-\frac{1}{1-r}}
\]
with
\[
g:=\sqrt[3]{\frac{(3-r)^2r^2}{4(1-r)^2}t_2^2\Big(-108x+12\sqrt{3}\sqrt{27x^2
-8\dfrac{(1-r)(2-r)^3}{(3-r)r}\dfrac{t_1^3}{t_2}}\Big) },
\]
and for $n=-2$, $C=0$
\[
v=\left(\frac{g(x,t_1,t_2)}{6(2-r)t_1}+\frac{2x^2}{3(2-r)t_1g(x,t_1,t_2)}-\frac{x}{3(2-r)t_1}\right)^{\frac{4}{1-r}}
\]
with
\begin{multline*}
g:=\left( -8x^3+
108\frac{(2-r)^2(3-r)(5-r)}{3(1-r)}t_1^2t_2\right.\\\left.+12\sqrt{3}(2-r)t_1\sqrt{\frac{(3-r)(5-r)}{3(1-r)}t_2\Big(
27\frac{(2-r)^2(3-r)(5-r)}{ 3(1-r)}t_1^2t_2-4x^3\Big)}
\right)^{\frac{1}{3}}.
\end{multline*}

\subsubsection{2-component reduction}

For $N=2$ we take the solution of \eqref{compatibility.rp.0.2.1}
and \eqref{compatibility.rp.0.3.1} given by
\[
\tilde \rho_1=-\tilde \rho_2=-\frac{2}{U_1-U_2},\quad \tilde
\pi_1=-\tilde\pi_2=\frac{1}{4(U_1-U_2)}.
\]
We need to handle \eqref{tilde-diag} which in this case reads
\begin{gather}
  \label{tilde-diag}
\frac{\partial \tilde f_1}{\partial U_2}=\frac{\partial \tilde
f_2}{\partial U_1}=-\frac{3-2r}{2(U_1-U_2)}(\tilde f_1-\tilde
f_2).
\end{gather}
which is equivalent to
\[
f_1=\frac{\partial \Phi}{\partial U_1},\quad
f_2=\frac{\partial\Phi}{\partial U_2},
\]
with
\[
\frac{\partial^2 \Phi}{\partial U_1\partial
U_2}+\frac{2r-3}{2(U_1-U_2)}\Big(\frac{\partial \Phi}{\partial
U_1}-\frac{\partial \Phi}{\partial U_2}\Big)=0.
\]
The method of separation of variables leads to the following
solutions:
\[
\Phi=(U_1-U_2)^{r-1}\big(A J_{1-r}(k (U_1-U_2))+B N_{1-r}(k
(U_1-U_2))\big)\big(C\cos(k (U_1+U_2))+D\sin(k (U_1+U_2))\big)
\]
and also
\[
\Phi=\begin{cases}   (A+B(U_1-U_2)^{2r-2})(C+D(U_1+U_2)), &r\neq
1,\\
(A+B\log(U_1-U_2))(C+D(U_1+U_2))& r=1.
\end{cases}
\]
From \eqref{relations.dym} we derive
\[
v=(U_1-U_2)^2,\quad v_0=\frac{U_1+U_2}{2},\quad
v_1=\frac{1}{16}+\frac{1}{(U_1-U_2)^2},
\]
observe that with this reduction we have
$v_1=\dfrac{1}{16}+\dfrac{1}{v}$. The hodograph system
\eqref{tilde.hodograph} is
\begin{align*}
  v^{2-r}\Big((3-r)\Big(\frac{1}{16}+\frac{2-r}{8}\frac{U_+}{U_-}\Big)t_2
  +\frac{2-r}{8}\frac{1}{U_-}t_1 \Big)+x+f_1&=0,\\
v^{2-r}\Big((3-r)\Big(\frac{1}{16}-\frac{2-r}{8}\frac{U_+}{U_-}\Big)t_2
  -\frac{2-r}{8}\frac{1}{U_-}t_1 \Big)+x+f_2&=0,
\end{align*}
which is equivalent to
\begin{align*}
\frac{3-r}{8}v^{2-r}t_2
 +2x+f_1+f_2&=0,\\
\frac{2-r}{4}\frac{v^{2-r}}{U_-}((3-r)U_+t_2+t_1) +f_1-f_2&=0.
\end{align*}
When $\Phi=0$ we get the solution
\begin{align*}
v&=\Big(-\frac{16}{3-r}\frac{x}{t_2}\Big)^{\frac{1}{2-r}},\\
v_0&=-\frac{1}{3-r}\frac{t_1}{t_2},\\v_1&=\frac{1}{16}+
\Big(-\frac{3-r}{16}\frac{t_2}{x}\Big)^{\frac{1}{2-r}}.
\end{align*}
Finally, if we set $\Phi=U_-^{2r-2}(A+BU_+)$ so that
\begin{align*}
  f_1+f_2&=2BU_-^{2r-2},\\
  f_1-f_2&=2(2r-2)U_-^{2r-3}(A+BU_+).
\end{align*}
and the hodograph system is
\begin{align*}
\frac{3-r}{8}v^{2-r}t_2
 +2x+2BU_-^{2r-2}&=0,\\
v^{2-r}((3-r)U_+t_2+t_1)
+\frac{8(2r-2)}{2-r}U_-^{2r-2}(A+BU_+)&=0.
\end{align*}
Observe that the first equation determines $U_-$ (or $v$), and
introducing this information in the second we get $U_+$ (or
$v_0$).

\section*{Acknowledgements}

The author is in debt with Luis Mart{\'\i}nez Alonso and Elena Medina,
coauthors of \cite{otros}, paper on which the present one is
partially based. Partial economical support from Direcci\'{o}n
General de Ense\~{n}anza Superior e Investigaci\'{o}n
Cient\'{\i}fica n$^{\mbox{\scriptsize \underline{o}}}$
BFM2002-01607 is also acknowledge.

\end{document}